\documentclass{aa}  

\usepackage{soul}
\usepackage{color}
\usepackage{amsmath}
\usepackage{multirow}
\usepackage{longtable}
\usepackage{threeparttable}
\usepackage{graphicx}
\usepackage{txfonts}
\begin{document}

   \title{Flaring activities of fast rotating stars have a solar-like latitudinal distribution}


   \author{Huiqin Yang \inst{1,2} \and Xin Cheng\inst{3}\and Jifeng Liu\inst{1,2,4}  \and Shuai Liu\inst{1} \and Zhanhao Zhao\inst{3}\and Guiping Zhou\inst{1,4,5} \and Yijun Hou\inst{1,4,5} \and Changliang Gao\inst{1}\and  Zexi Niu\inst{1,4}
          }

   \institute{Key Laboratory of Optical Astronomy, National
              Astronomical Observatories, Chinese Academy of Sciences, Beijing 100101, China\\
              \email{yhq@nao.cas.cn}\\ 
              \and
             Institute for Frontiers in Astronomy and Astrophysics, Beijing Normal University, Beijing 102206, China\\ 
              \and
              School of Astronomy and Space Science, Nanjing University, Nanjing 210093, China\\
              \and   
              School of Astronomy and Space Sciences, University of Chinese Academy of Sciences, Beijing 100049, China\\         
              \and
                State Key Laboratory of Solar Activity and Space Weather, Beijing 100190, China
             }


 
  \abstract
   {The dynamo theory has always been one of the biggest mysteries in stellar physics. One key reason for its uncertainty is poor knowledge of the dynamo process on stars other than the Sun. The most important observed feature of the solar dynamo is that active regions only appear at low latitudes, which provides a crucial constraint to the dynamo theory, while with Doppler imaging, the current technique to spatially resolve the stellar hemisphere, it is difficult to distinguish the equatorial region. As a consequence, the latitudinal distribution of active regions (LDAR) of stars is ambiguous and controversial, mainly due to the limitations of the current technique for spatially resolving the stellar surface. }
   {Fast rotating stars, which are young and active, are thought to operate with a different dynamo process than the Sun. We study their LDAR and compare them with the Sun to reveal the underlying dynamo process.}
   {Flares are drastic and observational activity events that occur in active regions. Here, we propose a new method for studying how the apparent flaring activity varies with respect to the inclination to determine the LDAR of fast rotating stars.}
   {We find that the LDAR of fast rotating stars is consistent with that of the Sun, contrary to expectations. Our results provide a crucial constraint to the stellar dynamo, indicating that the solar-like dynamo also applies to fast rotating stars, and even spans different stages of their evolution. }
   {}


   \maketitle
%

\section{Introduction}
One of the biggest mysteries in stellar physics is the dynamo theory, which explains how a star produces and sustains its magnetic fields \citep{Charb2020,Hath2015}. Until now, the dynamo theory had only been well tested through observations of the Sun, which has been regarded as an extraordinary laboratory for stellar physics. Nevertheless, the Sun is only one particular sample and therefore the constraints for dynamo theory are still very limited to date \citep{Charb2020,Choud2017}. Fortunately, this issue can be addressed by studying the dynamo process of stars and comparing them with the Sun, that is, through the solar--stellar connection method \citep{Brun2017}. However, the biggest obstacle to investigating the dynamo process of other stars is that they are spatially unresolved, the result being that the location information of the dynamo process is difficult to determine.

On the Sun, the most prominent feature of the dynamo process is that sunspots and active regions begin to appear at a latitude of around $30^{\circ}$  at the beginning of an activity cycle and propagate toward the equator as the activity cycle progresses \citep{Charb2020,Hath2015,Nandy2002}. This feature provides us with the latitudinal distribution of active regions (LDAR) of the Sun and provides a crucial constraint to the solar dynamo. The technique of Doppler imaging, which is used to reconstruct images of the stellar surface on other stars, does not clearly distinguish starspots near the equatorial region \citep{Rice2002,Berdy2005,Pisk1990,Roett2016}. The image can also be affected by stellar activities, resulting in spurious results \citep{Berdy2005,Strass2002,Bruls1998}. This means that the LDAR of other stars can be ambiguous and controversial \citep{Berdy2005}. 


Flares are a kind of drastic energy release event on the Sun driven by magnetic reconnection \citep{Hath2015}. Their locations can be used to locate where active regions appear. Assuming that flares on other stars originate from starspots, the locations of starspots can also be determined. The inclination (the angle of the stellar spin axis with respect to the observer’s line of sight) determines what latitude range of a star can be observed and which latitude makes the greatest contribution to the observation. By combining flares and inclinations of fast rotating stars that have similar intrinsic activity levels, we propose that the information on the LDAR for spatially unresolved stars can be revealed by the variation in the apparent flaring activity with the inclination.

\section{Data and method}
\subsection{Sample selection}

In this study, we combined the flaring activities of stars and their corresponding inclination to reveal information on the LDAR. This requires three observational quantities: flare, rotation, and projected rotation velocity($v$sin$i$), which are provided by the data from the {\it Kepler} mission and the Apache Point Observatory Galactic Evolution Experiment (APOGEE)-2 spectrograph (R $\sim 22,000$) of Sloan Digital Sky Survey (SDSS).

First, we selected common stars from the catalog of flaring stars \citep{Yang2019} and the catalog of rotation period \citep{MC2014} from the {\it Kepler} mission, which gave a total of $\sim 2200$ stars. We then crossmatched them with the catalog data from the APOGEE DR17\citep{Abdu2022}. About 240 stars remained. To exclude poor-quality data, we removed $\sim 20$ stars that have the following SDSS quality flags: (ASPCAPFLAGS): VSINI\_BAD,  SN\_BAD, STAR\_BAD,  CHI2\_BAD,  VMICRO\_BAD;  (STARFLAGS): BAD\_PIXELS, LOW\_SNR, or VERY\_BRIGHT\_NEIGHBOR. We also removed $\sim 20$ giants (log$g < 3.5$) whose spectrum broadening were dominated by macroturbulent velocity \citep{Holtzman2018}, and attained the final sample of 201 stars.

The final sample includes 21 RS Canum Venaticorum (RS CVn) binaries and 11 single-lined spectroscopic binaries that were identified as such by the non-single star and variability catalog of \textit{Gaia} DR3 \citep{Eyer2023}. We also identified six double-lined spectroscopic binaries (SB2s) in our sample (see Sect.~\ref{subsec:vsini}). Binaries are plotted in Fig. 1 with triangles, but are not included in the further analysis.

\subsection{Flaring activities of stars and the Sun}

Over four years of observation, the {\it Kepler} mission has provided continuous white-light curves of 0.2 million stars with unprecedented precision. About 3400 flaring stars produced 0.16 million flaring events \citep{Yang2019}. For each star, the flaring activity $R_{\rm flare}$ is defined as the ratio of the total flare energy to the total energy a star emitted during the observation \citep{Yang2017}:
\begin{equation}\label{eq_fa}
 R_{\rm flare}=\frac{\sum E_{\rm flare}}{\int L_{\rm bol}dt} = \frac{\tilde{L}_{\rm flare}}{L_{\rm bol}}.
\end{equation}

Here, $\sum E_{\rm flare}$ is the sum of all detectable flare
energies during the whole observation and $L_{\rm bol}$ is the bolometric luminosity. The definition of $R_{\rm flare}$ is similar to the proxy of chromospheric activity $R'_{\rm HK}$ \citep{Noyes1984} and coronal activity $R_{\rm X}$ \citep{Wright2011} and is a normalized quantity that has removed the influence of stellar luminosity. The left panels of Fig. 1 show an example of flaring stars with different activity levels, which can be visually distinguished from the frequency and size of flares.

\begin{figure*}
\includegraphics[width=1.0\textwidth]{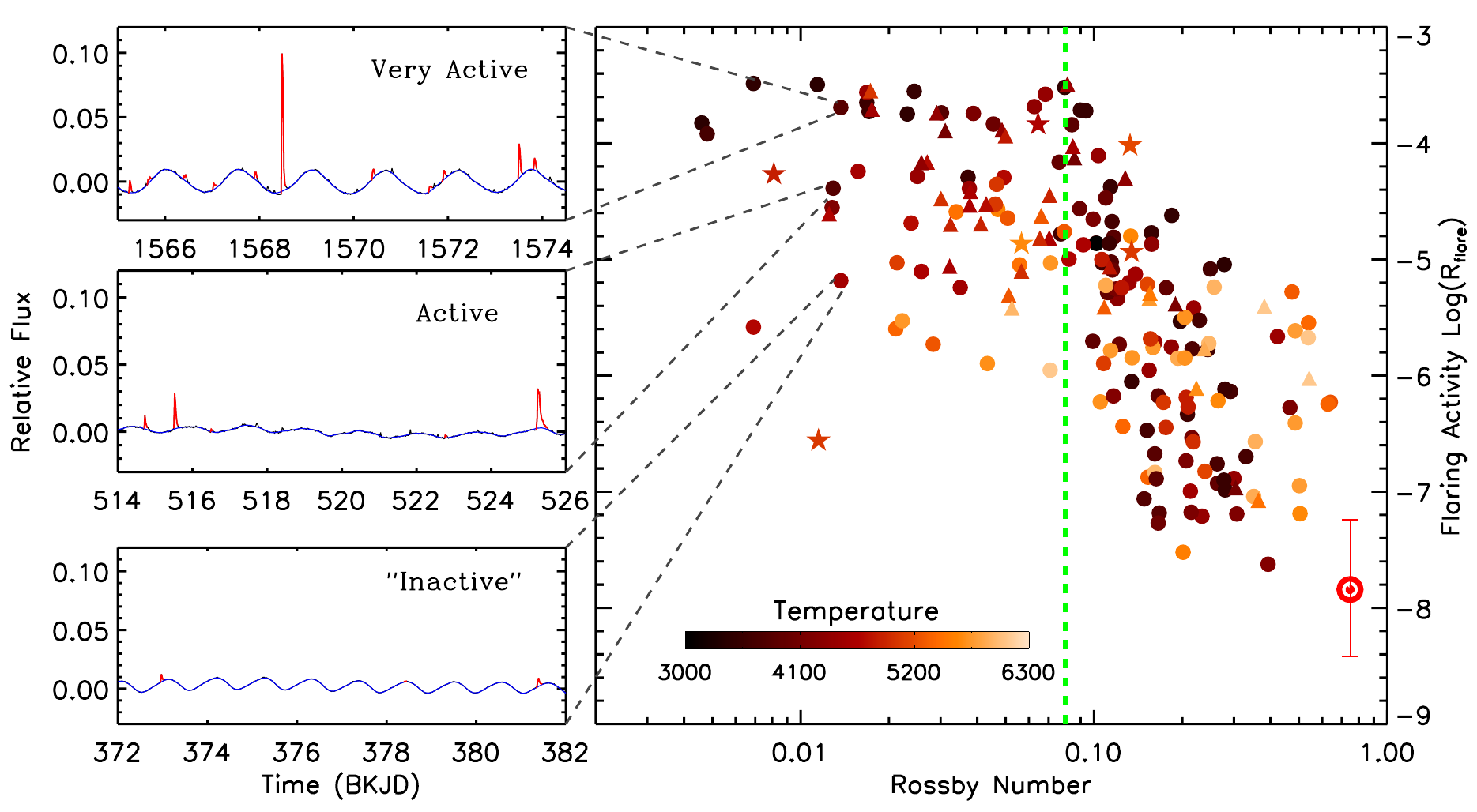}
\caption{Left panels: Example of flaring stars with different activity levels. Red curves denote flares. Blue curves are fitted baselines. Black curves are the relative flux.  Right panel: Rotation--flaring activity relationship. The $x$ axis is the Rossby number (Ro; the ratio of the rotation period to the global convective turnover time), which has removed the influence of stellar mass. The shapes refer to the following star types: circle = dwarf; triangle =  binary; and five-pointed star = subgiant. The green line (Ro = 0.08) separates the saturated and unsaturated regimes. The Sun is marked with an $\odot$ symbol.}
\label{fig1}
\end{figure*}
The right panel of Fig. 1 shows the rotation--activity relationship in terms of flaring activity, where the $x$ axis is the Rossby number (Ro; the ratio of the rotation period to global convective turnover time) instead of the rotation period. This canonical relationship has been validated by various activity proxies \citep{Noyes1984,Wright2011,Yang2019,Yang2017}, and it identifies two stages of stellar evolution: the saturated (faster, with higher activity) and unsaturated (slower, with declining activity) regimes. In the saturated regime, the stellar activity of those fast rotating stars becomes saturated, such that they are thought to have the same intrinsic activity level.
\begin{figure*}
\includegraphics[width=0.32\textwidth]{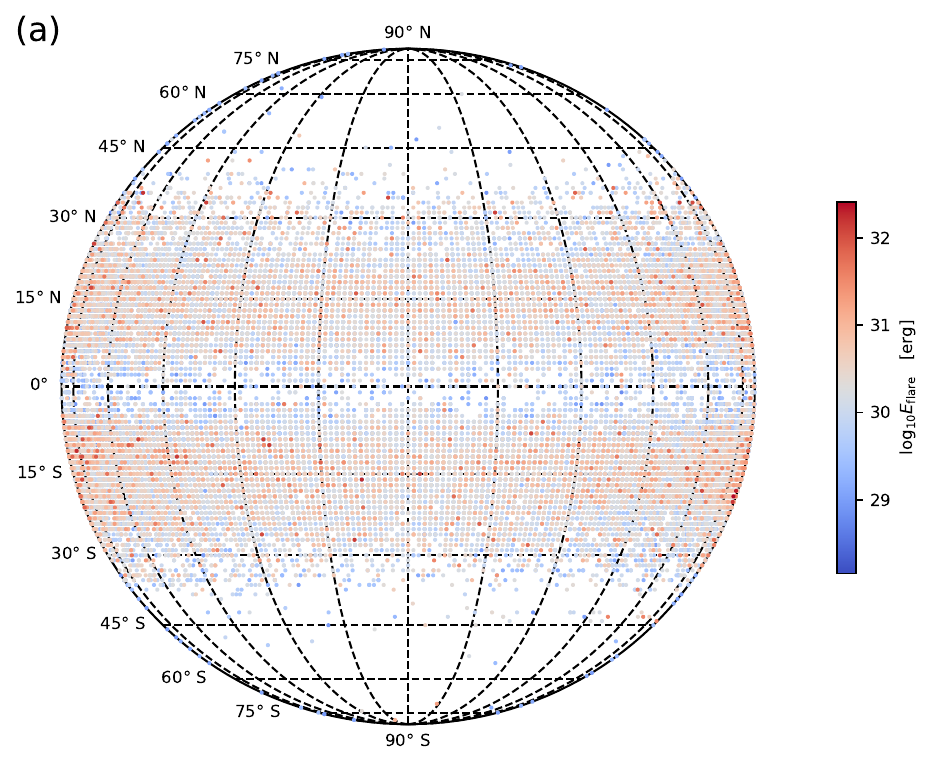}
\includegraphics[width=0.32\textwidth]{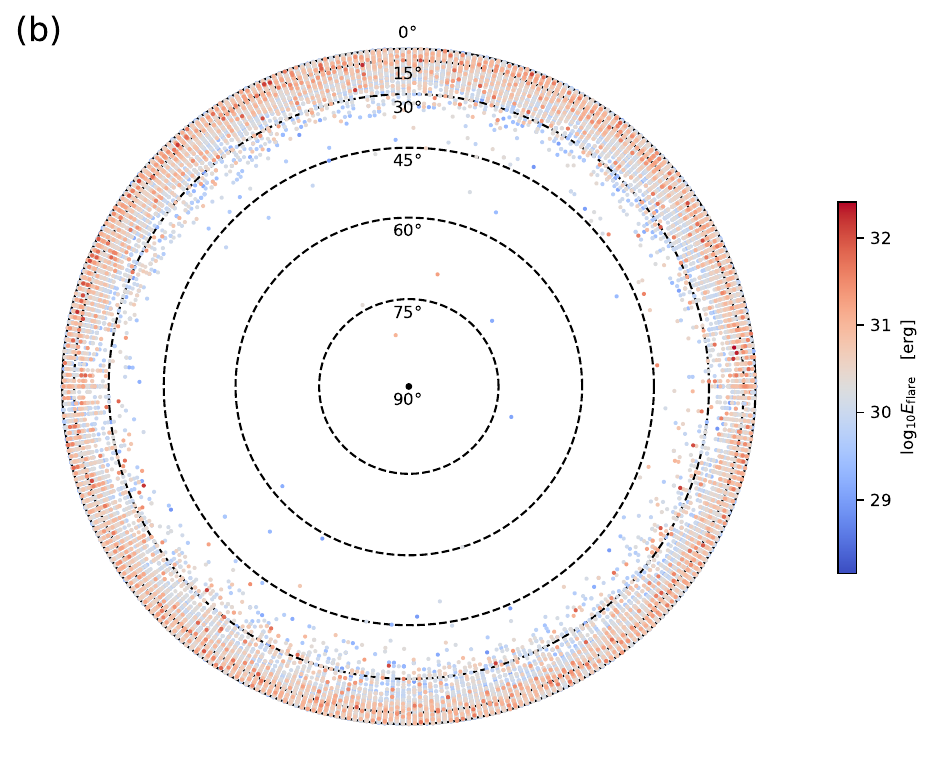}
\includegraphics[width=0.32\textwidth]{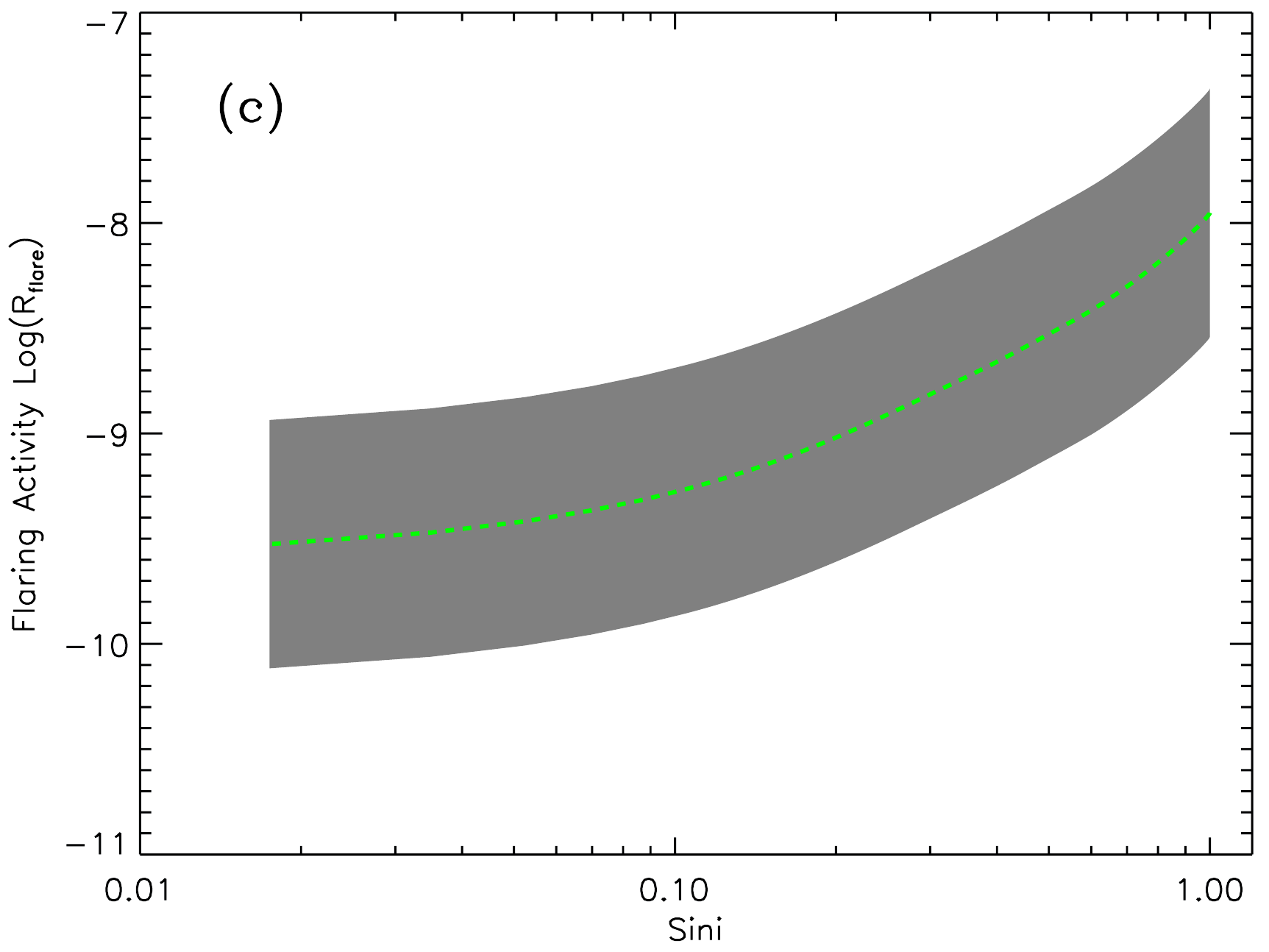}
\caption{Panel (a): About 38000 solar flares from 1975 to 2017  superimposed on the solar hemisphere that is observed from edge-on. Panel (b): Same as (a) but observed from pole-on. Panel (c): Observed flaring activity of the Sun varies with the inclination. The shaded region represents the uncertainty caused by the maximum and minimum of the solar cycle.}
\label{fig2}
\end{figure*}

\begin{figure*}
\includegraphics[width=0.32\textwidth]{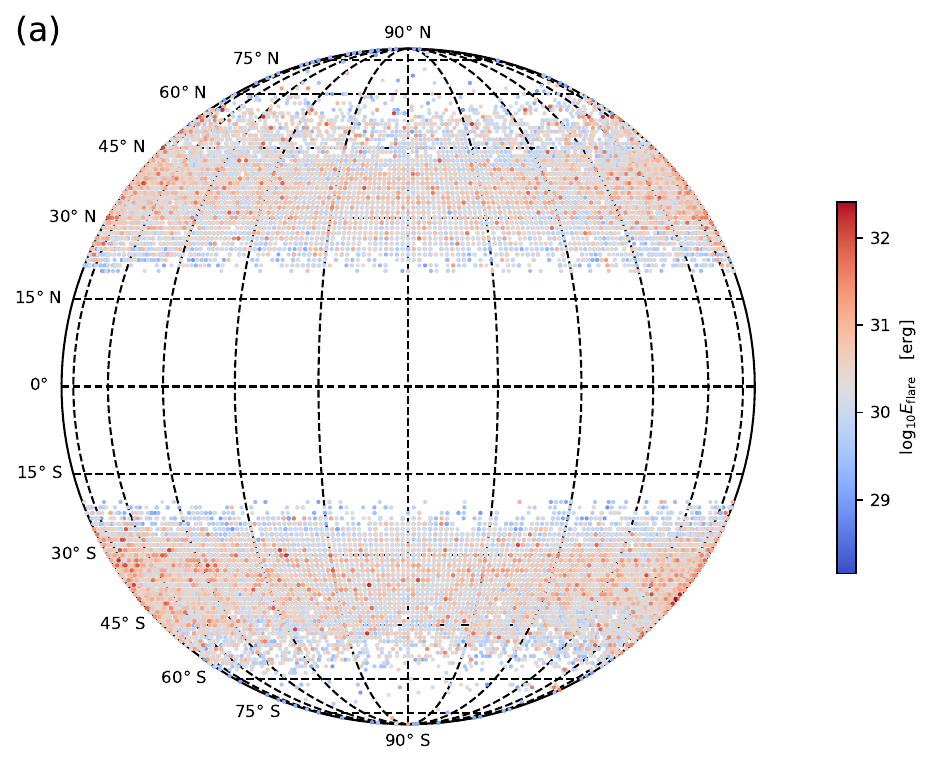}
\includegraphics[width=0.32\textwidth]{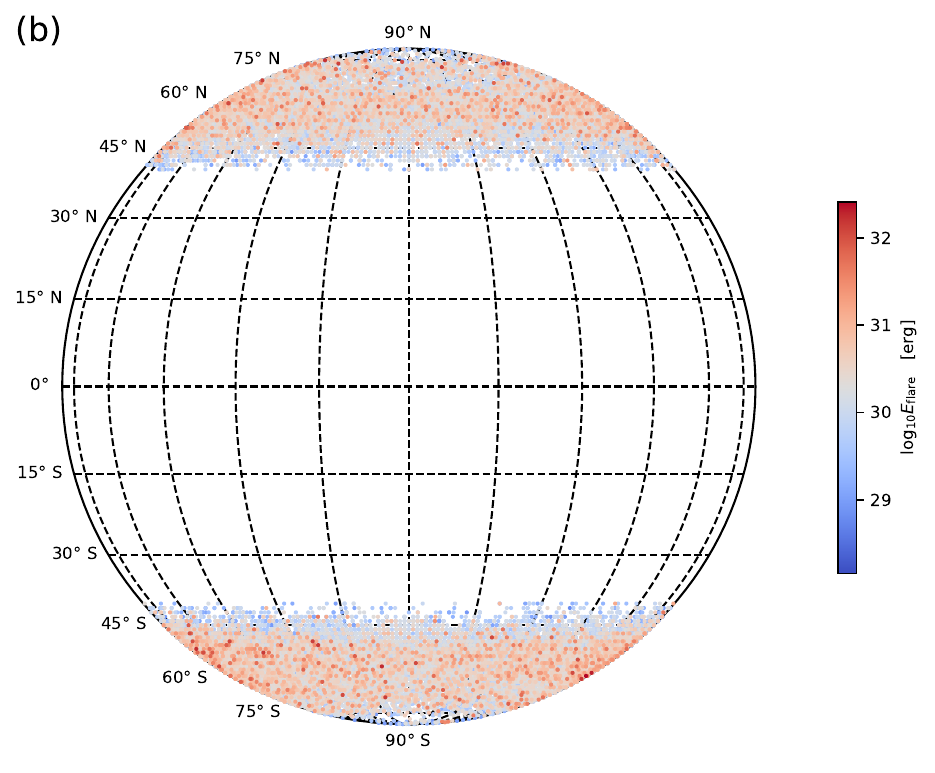}
\includegraphics[width=0.32\textwidth]{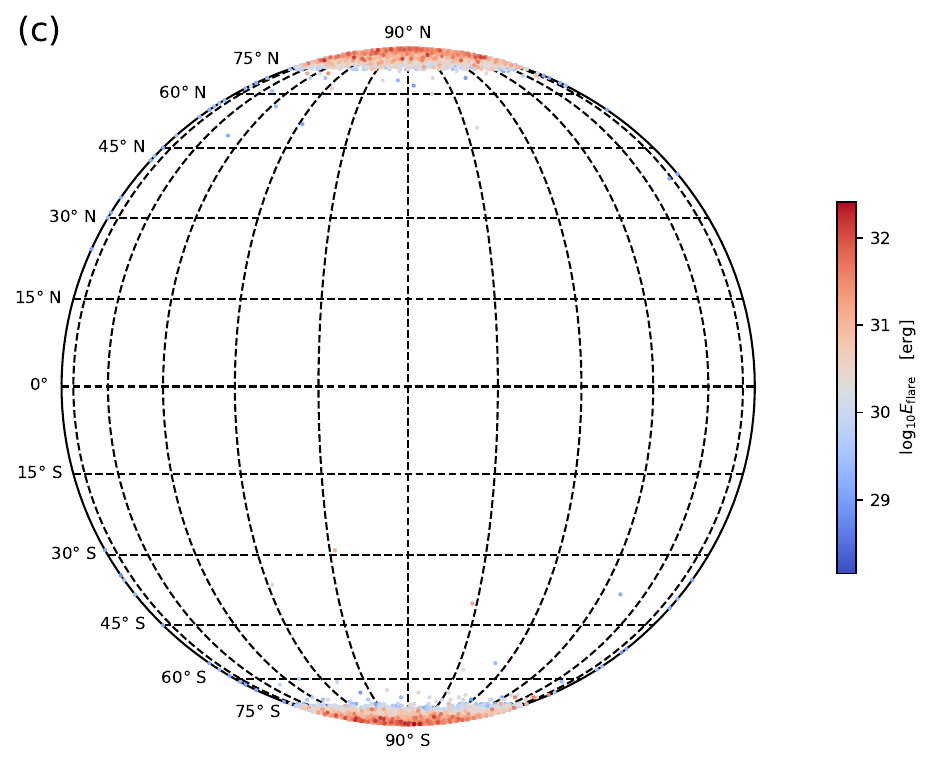}
\includegraphics[width=0.32\textwidth]{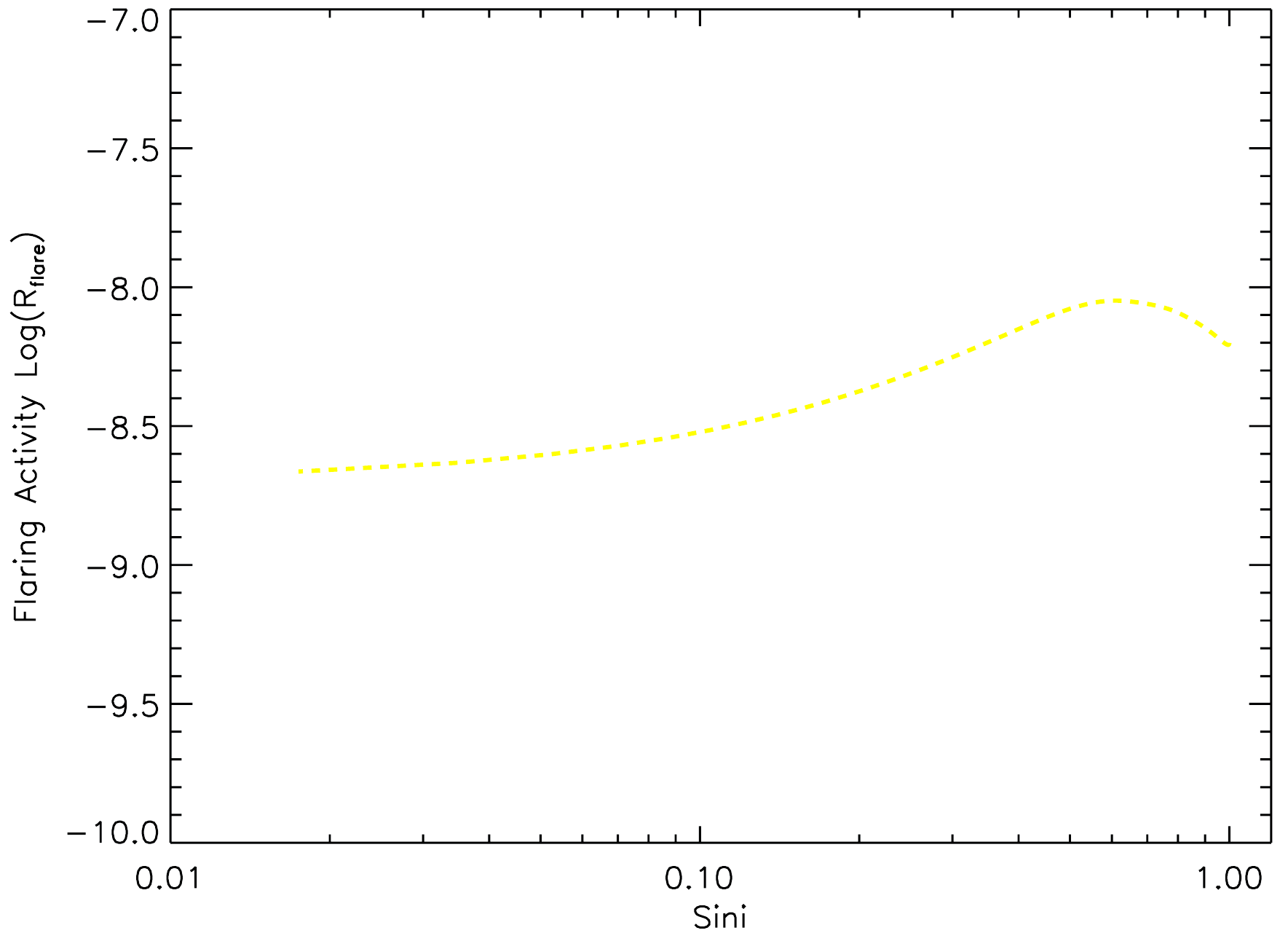}
\includegraphics[width=0.32\textwidth]{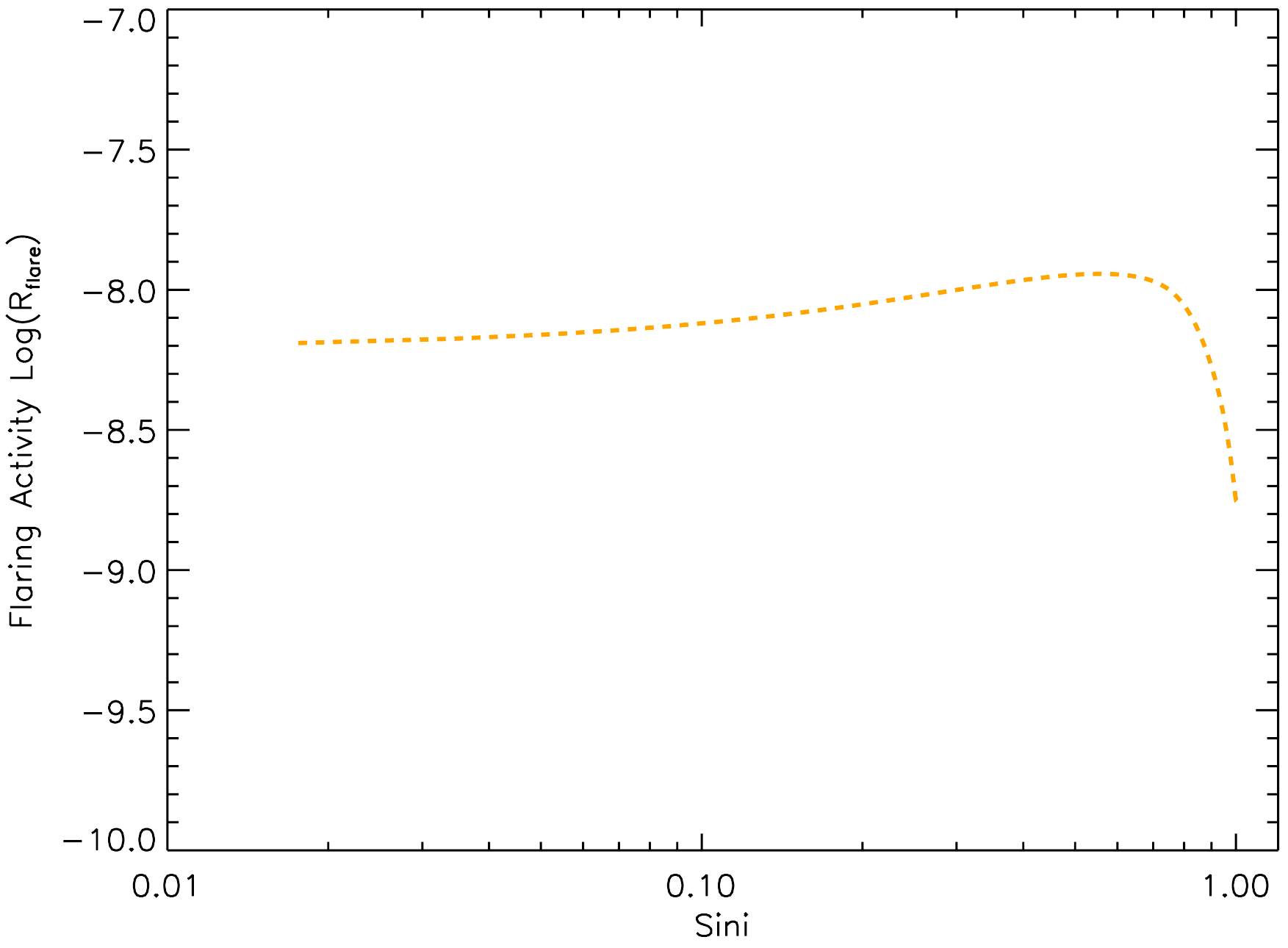}
\includegraphics[width=0.32\textwidth]{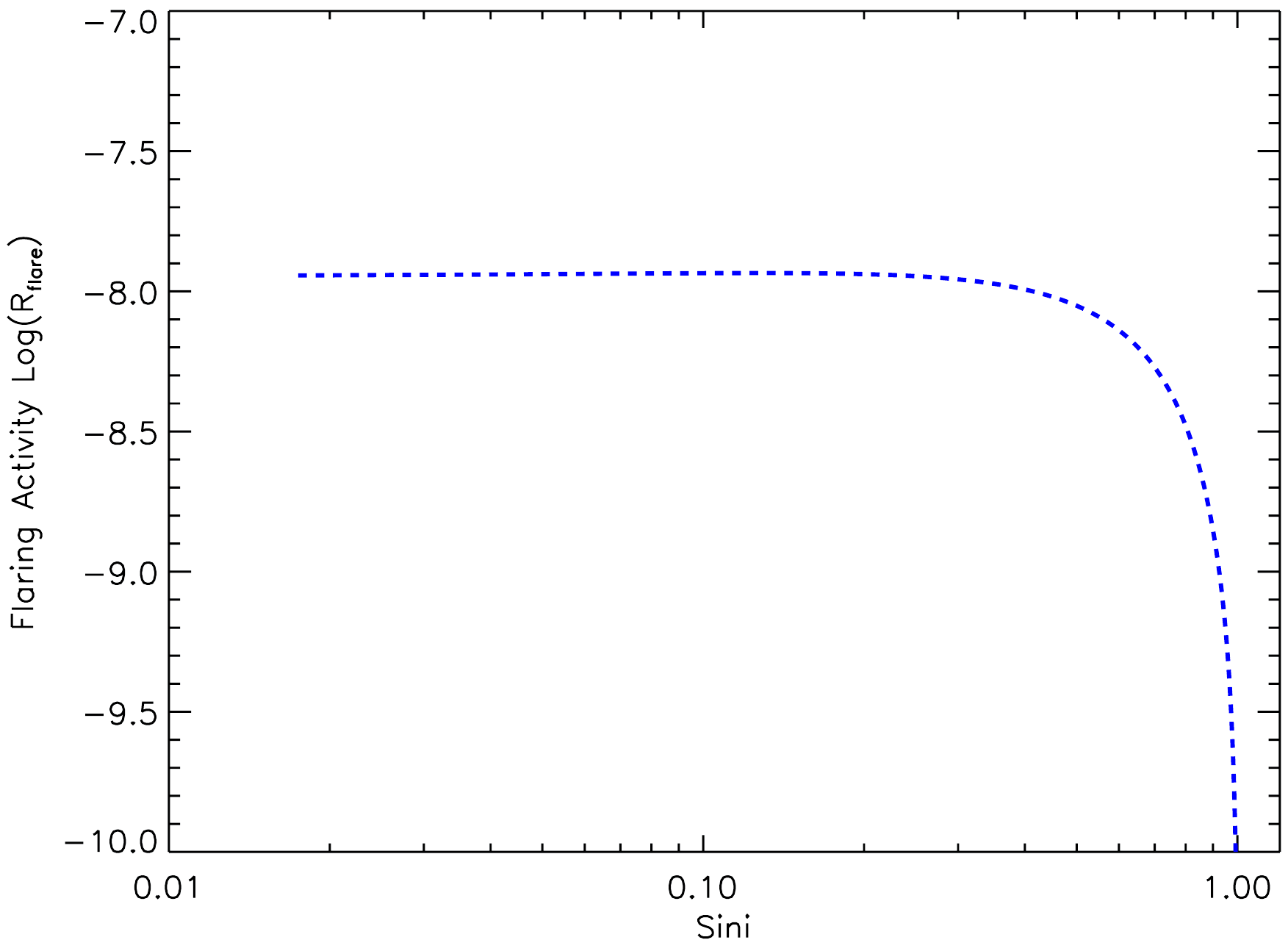}

\caption{Flaring activity vs. inclination for different LDARs. The top panels show the flaring region at mid latitude, high latitude, and the polar region. The different flaring regions are created by increasing the latitude of solar flares by 20$^{\circ}$, 40$^{\circ}$, and 70$^{\circ}$, respectively. The bottom panels show the relation between the flaring activity and inclination for the corresponding flaring regions.}
\label{fig_simu}
\end{figure*}
However, given that the Kepler mission provides curves of the white-light band, the flaring location  \citep{Wata2013} and the limb darkening effect  \citep{Claret2011} will be the main factors that influence the observed flare energy and number, and in turn determine the apparent flaring activity. Figure 2 shows the solar surface on which $\sim 38000$ solar flares from 1975 to 2017 are superimposed. We simulated an observation of the Sun as a star by the {\it Kepler} mission from equator-on (Fig. 2a inclination $i \approx 90^\circ$) to pole-on (Fig. 2b $i \approx 0^\circ$; see Sect.~\ref{subsec:simulation} for details of the simulation). The variation in inclination determines whether and where each flare is observed on the hemisphere. This results in a dramatic rise in the apparent flaring activity, with an increase in inclination for the latitudinal distribution of solar flares (Fig. 2c).

For comparison, we created three latitudinal distributions of solar flares by taking the equator as the axis of symmetry and increasing the latitude of flares by 20$^{\circ}$, 40$^{\circ}$, and 70$^{\circ}$, respectively (Fig.~\ref{fig_simu}). These three distributions correspond to the active regions at mid-latitude, high-latitude, and the polar regions, respectively. Equally, we simulated observations of them from different inclinations and obtained three distinct relationships between flaring activity and inclination (Fig.~\ref{fig_simu}).
\subsubsection{Flare detection and energy estimate}

The flare detection is similar to previous studies \citep{Yang2017,Yang2018,Yang2019}. We briefly summarize the method here. (1) We used a smoothing filter based on the spline algorithm to fit the quiescent flux or the baseline, where an iterative $\sigma$-clipping approach was applied to remove all outliers. (2) After detrending the baseline, we applied several criteria to identify flare candidates, including the amplitude and duration (at least three continuous points higher than 3$\sigma$), the profile (the decay phase should be longer than the rise phase), and a check of the false-positive signal (no break points within 3 hr, target pixel file check, neighboring star check). (3) All of the flares were manually checked, especially those whose duration was comparable to the rotation period. We reconstructed the quiescent fluxes using the neighboring data points of the rotation period. The left panels of Fig. 1 show examples of flare detection.

We estimated the flare energy by assuming a flare radiated as a blackbody with an effective temperature of 9000 K per unit area \citep{Kretz2011}, and the flare area was estimated based on the flare amplitude, the stellar radius, the stellar effective temperature, and the Kepler
response function  \citep{Shibayama2013,Yang2017,Yang2019}. The uncertainty of the flare energy is $\sim 60\%$ \citep{Shibayama2013}.

\subsubsection{Flaring activity of the Sun}\label{subsec:sunfa}
In order to quantify the flaring activity of the Sun ($R_{\rm sun}$), we first used the data of Geostationary Operational Environmental Satellites (GOES) soft X-ray (SXR) peak flux ($F_{\rm SXR}$) to estimate the bolometric flare energy ($E_{\rm flare}$). The comparison between the $F_{\rm SXR}$ of SXR and the $E_{\rm flare}$ \citep{Kretz2011,Name2017} can provide an empirical relation:

\begin{equation}\label{eq_egyconvt}
\begin{aligned}
 {\rm log}(E_{\rm flare})&= a \cdot {\rm log}(F_{\rm SXR}) +b \\
 a&= 0.78^{+0.08}_{-0.08}\\
 b&=34.4^{+0.03}_{-0.03}. 
\end{aligned}
\end{equation}

\begin{figure}%
\includegraphics[width=0.5\textwidth]{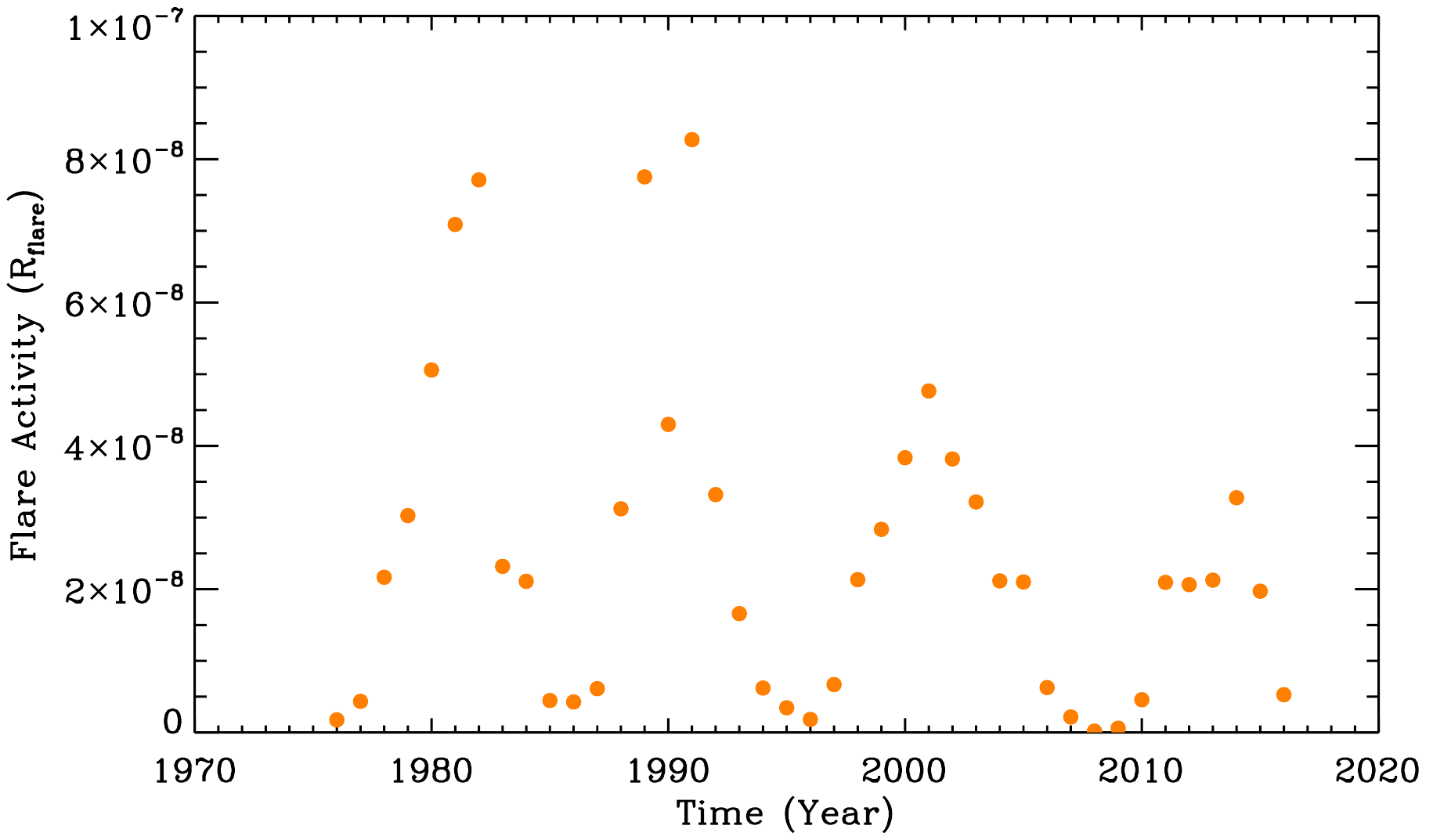}
\caption{Flaring activity of the Sun from 1975 to 2017. The solar flare energies are estimated based on the GOES SXR data (Eq.~\ref{eq_egyconvt}). The variation in the flaring activity shows an 11-year cycle.
}\label{fig_circle}
\end{figure}

The National Oceanic and Atmospheric Administration (NOAA) \footnote{Available online: URL https://www.ngdc.noaa.gov/stp/space-weather/solar-data/solar-features/solar-flares/x-rays/goes/xrs/} has continuously recorded GOES SXRs of $\sim 77000$ solar flares from 1975 to 2017. We converted their $F_{\rm SXR}$ to $E_{\rm flare}$ and calculated the flare activity, $R_{\rm flare}$ , of the Sun (Eq. ~\ref{eq_fa}) for each year (Fig.~\ref{fig_circle}). Figure~\ref{fig_circle} shows an 11-year cycle of solar flaring activity that is the same as sunspots and other activity proxies. 

We then adopted the mean flaring activity of 33 years (three flaring cycles from 1976 to 2008) and the standard deviation as the flaring activity of the Sun and its uncertainty respectively.  The location of the Sun is marked with an $\odot$ symbol in Fig.~\ref{fig1} and its value is 

\begin{equation}\label{eq_sunfa}
 {\rm log}(R_{\rm sun}) = -7.83 ^{+0.59}_{-0.59}.
\end{equation}

\subsection{Rossby number and the global convective turnover time}\label{subsec:tau_g}
The Rossby number is the ratio of rotation period to convective turnover time. The convective turnover time cannot be observed directly and instead has to be estimated using either empirical fits \citep{Noyes1984,Wright2011} or computations of theoretical models  \citep{Spada2017,Kim1996}. 

We used a grid of stellar evolutionary tracks from the Yale-Potsdam Stellar Isochrones (YaPSI)  \citep{Spada2017} to derive the global convective turnover time $ (\tau_g$ ) by finding their best match to the physical parameters of each star. Our method is similar to that of a previous study \citep{Lehtinen2020}. We first remapped the YaPSI tracks as a function of uniformly spaced ``equivalent evolutionary points"(EEPs) \citep{Spada2017}. We then constructed a series of synthetic tracks by linearly interpolating EEPs with steps of 0.02 $M_{\odot}$. We compared the $T_{\rm eff}$ and log$g$ of each star with tracks from $0.2 M_{\odot}$ to $3M_{\odot}$ and obtained $\tau_g$ from the closest point of all the tracks to the star. In this way, we could repeat the above procedures to match $\tau_g$ for different metallicities from -1.5 to 0.3. The final value of $\tau_g$ was obtained via linear interpolation in of the metallicity.

The traditional rotation--activity relationship usually uses the local convective turnover time $\tau_l$, but $\tau_g$ is preferred here because $\tau_l$ is obtained via empirical fits \citep{Noyes1984,Wright2011} that have hardly taken subgiants and giants into account. Our sample comprises subgiants that also conform to the rotation--activity relationship in terms of $\tau_g$  \citep{Lehtinen2020}. The relation between $\tau_l$ and $\tau_g$ has been thoroughly discussed \citep{Kim1996,Lehtinen2020,Spada2017,Landin2010,Mittag2018}. In mathematics,$\tau_g \simeq 2.5 \tau_l$ \citep{Lehtinen2020,Mittag2018,Kim1996,Landin2010}, which makes the critical Rossby number that separates the saturated and unsaturated regime  $\sim 0.1$.

\subsection{Inclination} \label{subsec:inclination}
The inclination of a star is the angle of the stellar spin axis with respect to the observer's line of sight. Its sine value can be derived through the spectro-photometric method, which is governed by the following equation  \citep{Jackson2010,Healy2020,Healy2023}:


\begin{equation}\label{eq_inclination}
{\rm sin}i = \frac{v {\rm sin} i \cdot P}{2\pi R}.
\end{equation}
Here,  $v$sin$i$ is the projected rotational velocity measured from high-resolution spectral broadening  \citep{Abdu2022}, $R$ is the stellar radius, which can be estimated from the isochrone fitting \citep{Somers2020}, and $P$ is the rotation period measured from light curve modulation \citep{MC2014}. 

\subsubsection{Projected rotation velocity} \label{subsec:vsini}
\begin{figure}%
\includegraphics[width=0.45\textwidth]{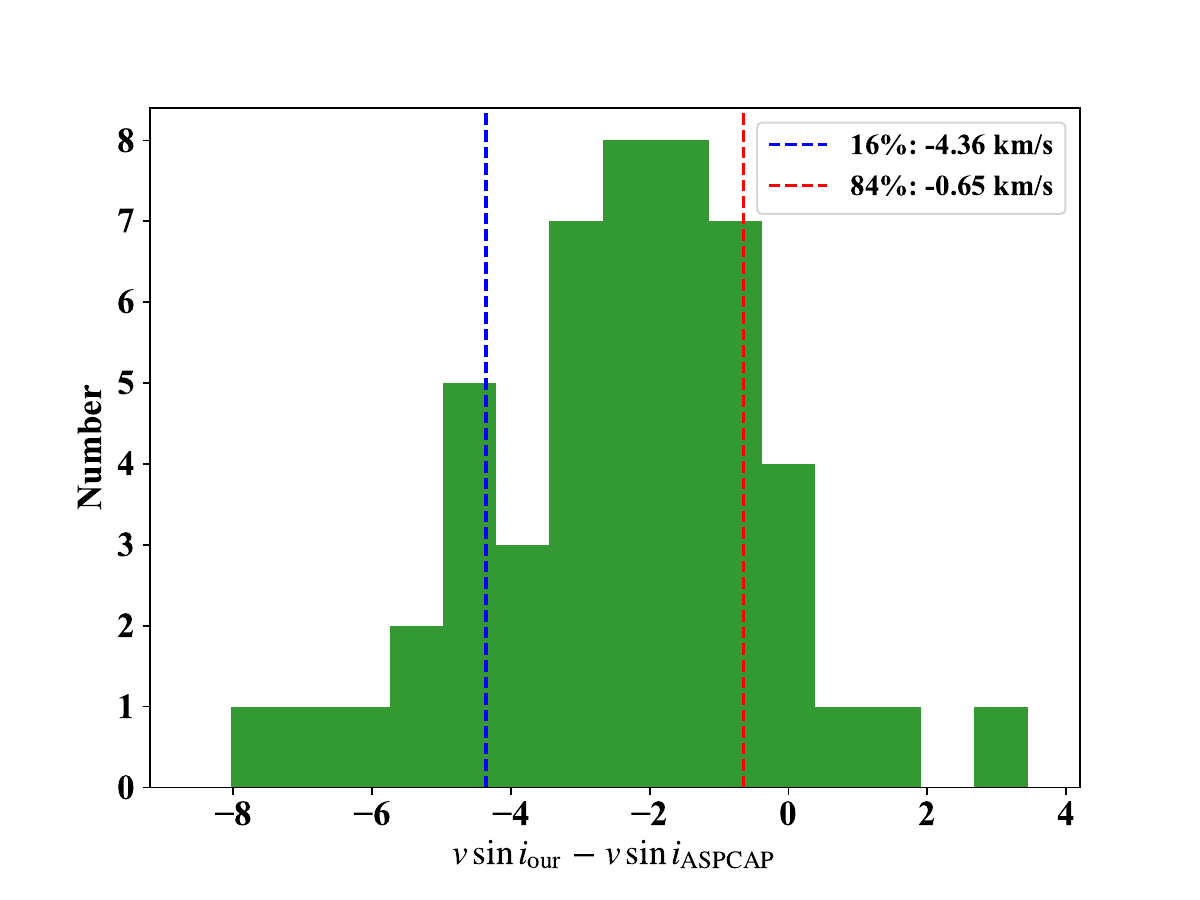}
\caption{Comparison of the $v$sin$i$ from our measurements and that of ASPCAP  for stars of  $v$sin$i > 10 {\rm ~km~s^{-1}}$. The vertical dashed blue and red  lines denote the 16th and 84th percentiles of the distribution, respectively.
}\label{fig_vsini_comp}
\end{figure}
We measured the $v$sin$i$ of stars in our sample using spectra from APOGEE DR17 \citep{Abdu2022}, which covered the Kepler field. We first generated theoretical templates using 1D local thermodynamic equilibrium model atmospheres with a radiative and convective scheme (MARCS) stellar atmosphere models. These templates were convolved with a Gaussian kernel (R $\approx 13 {\rm ~km~s^{-1}}$) to simulate instrumental broadening. We then convolved these nonrotating templates with varying profiles of rotation to determine the best match of $v$sin$i$ to the observation. In the matching process, we focused on five Fe I lines (15207.54, 15294.56, 15621.65, 15980.73, and 16486.69 {\AA}) that exhibit minimal blending. The final result of $v$sin$i$ is the mean value calculated from these Fe I lines.

In order to validate our measurements, we compared them with the values of $v$sin$i$ provided by the Apogee Stellar Parameter and Chemical Abundances Pipeline (ASPCAP). Our results generally show good agreement with ASPCAP’s findings, though there is a systematic offset of $\sim 3{\rm ~km~s^{-1}}$(Fig.~\ref{fig_vsini_comp}). We plotted fitting lines of five stars for which the difference in $v$sin$i$ between our measurement and that of ASPCAP is larger than $ 5 {\rm ~km~s^{-1}}$ (Fig.~\ref{fig_vsini_examp2}). We note that no matter which measurement is used, it will not change the results. During our measurement process, we found six SB2s that seriously biased the ASPCAP measurements (Fig.~\ref{fig_vsini_examp}). We plot them in Fig. 1 with triangles, but did not include them in the further analysis.

The mechanisms of spectroscopic broadening are mainly from rotation and macro-turbulence. The solar macro-turbulent velocity is $\sim 3~{\rm km~s^{-1}}$ and that of cooler stars is slightly lower  \citep{Doyle2014}.  We therefore assumed the uncertainties of $v$sin$i$ to be 3 ${\rm ~km~s^{-1}}$ or $10\%$, whichever was greater  \citep{Tayar2015}. Due to the resolution of APOGEE, the variance of its $v$sin$i$ rapidly increases below $10{\rm ~km~s^{-1}}$  \citep{Simonian2020}. For stars of $v$sin$i < 10 {\rm ~km~s^{-1}}$, we used the ASPCAP measurements  and adopted a conservative policy by setting the lower error bar and upper error bar to be $1{\rm ~km~s^{-1}}$ and $10{\rm ~km~s^{-1}}$ respectively, although the detection limit of APOGEE is $\sim 5{\rm ~km~s^{-1}}$ \citep{Desh2013,Gil2018}.

\subsubsection{Stellar radius} \label{subsec:radii}

There is a widespread discrepancy between the radii of young, active stars and model predictions, which is the so-called radius inflation. This discrepancy is quite serious for fast rotating stars, being up to $20\%$  \citep{Jackson2014,Somers2017,Jaehnig2019}. This is attributed to the fact that current stellar models have a simple treatment of the surface convection and do not account for the influence of magnetic activity on stellar parameters \citep{Spada2013,Somers2020}. For example, starspots inhibit convection, which reduces the effective temperature of a star. The temperature cooling will result in a smaller radius in the isochrone fitting.

In order to estimate the radii of our rapidly rotating, active stars, we used the stellar parameters of tracks with starspots (SPOTS) \citep{Somers2020} to fit the effective temperature and log$g$ of the APOGEE parameters. The SPOTS models include the structural effect of starspots in solar-metallicity evolutionary tracks, ranging from spotless to a surface-covering fraction ($f_{\rm spot}$) of $85\%$.   \citet{Fang2016} found that in the saturation regime, the covering fraction of a spot is $\sim 30\%-40\%$ for G-type stars and $\sim 40\%-50\%$ for K- and M-type stars. Therefore, we adopted the SPOTS tracks of $f_{\rm spot}=0.34$ for G-type stars ($T_{\rm eff} > 5000 K$) and $f_{\rm spot}=0.51$ for K- and M-type stars ($T_{\rm eff} < 5000 K$). We also carried out isochrone fitting through the traditional models of PAdova-TRieste Stellar Evolution Code (PARSEC) \citep{Chen2014} and YaPSI \citep{Spada2017} and compared them with the SPOT-derived radii. We found that traditional models underestimated the radii of rapid rotating stars by, on average, $\sim 9\%$ for G-type stars and by $\sim 11\%$ for K- and M-type stars. The comparison is consistent with previous studies on stars in open clusters \citep{Somers2017,Jaehnig2019}. 

\subsubsection{Determination of inclination} \label{subsec:PPD}
Given that we determined sin$i$ through Eq.~\ref{eq_inclination}, which involved a comparison of two correlated parameters (the calculated $v$ and measured $v$sin$i)$, we thus performed Bayesian inference and determined the posterior probability distribution (PPD) of sin$i$ for each star   \citep{Masuda2020}. 
We used the calculated $v$  and measured $v$sin$i$ and their uncertainties to define the likelihood function of $v$ and $v$sin$i$, both of which are assumed to follow a Gaussian distribution. We set uniform priors for sin$i$ and $v$ \citep{Masuda2020}. We applied Bayes' theorem to set up an integral calculating $p$(sin$i|D$) and directly computed the integral for the PPDs  \citep{Healy2020}. For each distribution, we adopted the median value as the sin$i$, with the 16th and 84th percentiles providing the uncertainties.

\subsubsection{Uncertainties of inclination}\label{subsec:uncertainty}
We summarize the factors that can cause uncertainties in the parameters in Eq.~\ref{eq_inclination} , which will result in an overestimate or underestimate of sin$i$. And we discuss their influence on our results below.

\textbf{Overestimate of sin$i$:} (1) Differential rotation will cause an overestimate of the rotation period, $P$, if it is a solar-like differential rotation (i.e., the pole rotates more slowly than the equator).  (2) $v$sin$i$ will be overestimated by turbulence and activity, which are serious factors for slow rotators. (3) $R$ is underestimated by the stellar model especially for fast rotators (radius inflation).

\textbf{Underestimate of sin$i$:} (1) Differential rotation will cause an underestimate of the rotation period, $P$, if it is the antisolar differential rotation (i.e., the pole rotates more quickly than the equator). (2) Differential rotation will cause an underestimate of $v$sin$i$ by a factor of $\sim 5-10\%$ in solar-like rotators  \citep{Hirano2014}.


We dealt with overestimation issues (2) and (3) by setting a proper uncertainty or taking account of the temperature cooling, while item (1) of overestimate and the underestimate of sin$i$ are more complicated. Firstly, a study of the asteroseismology of 40 solar-like stars shows that one-fourth of them have antisolar differential rotation  \citep{Benomar2018}, although the significance is not enough. Further analysis on the significance of their signals and theoretical expectation suggests that fast rotators are dominated by solar-like differential rotation. We thus assumed that stars of our sample have solar-like differential rotation \citep{Gastine2014}. The overestimated period and the underestimated $v$sin$i$ will act in opposite directions on the determined inclination. Their effects will partially cancel out and cause a mean systematic underestimate of sin$i$ by $5.7\%$  \citep{Healy2023}. We added this fractional error in quadrature to the upper error bar of our results.

\subsection{Observing the Sun as a star through the Kepler mission telescope}\label{subsec:simulation}
We simulated observations of the Sun through the Kepler mission telescope from different inclinations.
We  considered two factors that influence the observation of a flare. One is the visible fraction, $D$, of a latitude with respect to an inclination angle. It is governed by the following equation  \citep{Ilin2021}:

\begin{equation}
 D = \begin{cases} 1,& -{\rm tan}\theta \enspace {\rm tan}(\frac{\pi}{2}-i) < -1
 \cr \frac{1}{\pi} {\rm arccos} (-{\rm tan}\theta \enspace {\rm tan}(\frac{\pi}{2}-i)), & -1 \leq -{\rm tan}\theta \enspace {\rm tan}(\frac{\pi}{2}-i) \leq 1
 \cr 0, &-{\rm tan}\theta \enspace {\rm tan}(\frac{\pi}{2}-i) > 1 \end{cases}.
\end{equation}
Here, the inclination angle is $i$ and the latitude is $\theta$. This equation determines the probability that a flare is observed at a given latitude and inclination. We take it as the observed fraction of the flare energy at the latitude.

Another factor is the limb-darkening effect in the Kepler band \citep{Claret2011}. We adopted the quadratic function of the limb-darkening law:
\begin{equation}
\begin{aligned}
    \frac{I(\mu)}{I(1)}&=1-a(1-\mu)-b(1-\mu)^2 \\
    \mu&={\rm cos}(\phi)\\
    a&=0.40\\
    b&=0.26 
\end{aligned}
.\end{equation}
Here, $I$ is the observed intensity and $\phi$ is the angle between the line of sight and the normal to a given point on the stellar surface. The parameters $a$ and $b$ are the limb-darkening coefficients obtained under the condition that $T_{\rm eff} =5750 {\rm K}, {\rm log}g = 4.5,[{\rm M/H}] =0,$ and microturbulent velocity is 2 ${\rm km~s^{-1}}$  \citep{Claret2011}. We note that the influence of magnetic activity on the limb-darkening coefficients is less than $5\%$  \citep{Kost2024}, which is negligible in this study.

We compared the flare flux intensity $(F)$ and the flare duration ($\tau)$ of the solar and stellar flares  \citep{Name2017,Maehara2015} and obtained $\tau \sim F^{2.73}$. Given that the relation between the flare energy ($E)$ and the flare duration was $E \sim \tau ^{3}$ \citep{Name2017,Maehara2015}, we obtained $E \sim F^{8.19}$.  

Based on the above relations, we obtained the relation between the observed flare energy $(E_{\rm observed}$ ) and the intrinsic flare energy $(E)$:

\begin{equation}\label{eobs}
E_{\rm observed} = E \cdot D \cdot (1-a(1-\mu)-b(1-\mu)^2)^{8.19}
.\end{equation}
Using Eq. \ref{eobs}, we calculated the apparent flaring activity with decreasing inclination from equator-on to pole-on; the results are shown in Fig. 2c. We then increased the latitude of the solar flares by 20$^{\circ}$, 40$^{\circ}$, and 70$^{\circ}$ , respectively, to simulate LDAR in mid-latitude, high-latitude, and polar regions (Fig.~\ref{fig_simu}) and obtained three distinct relationships between the flaring activity and inclination.




\begin{figure}
\includegraphics[width=0.5\textwidth]{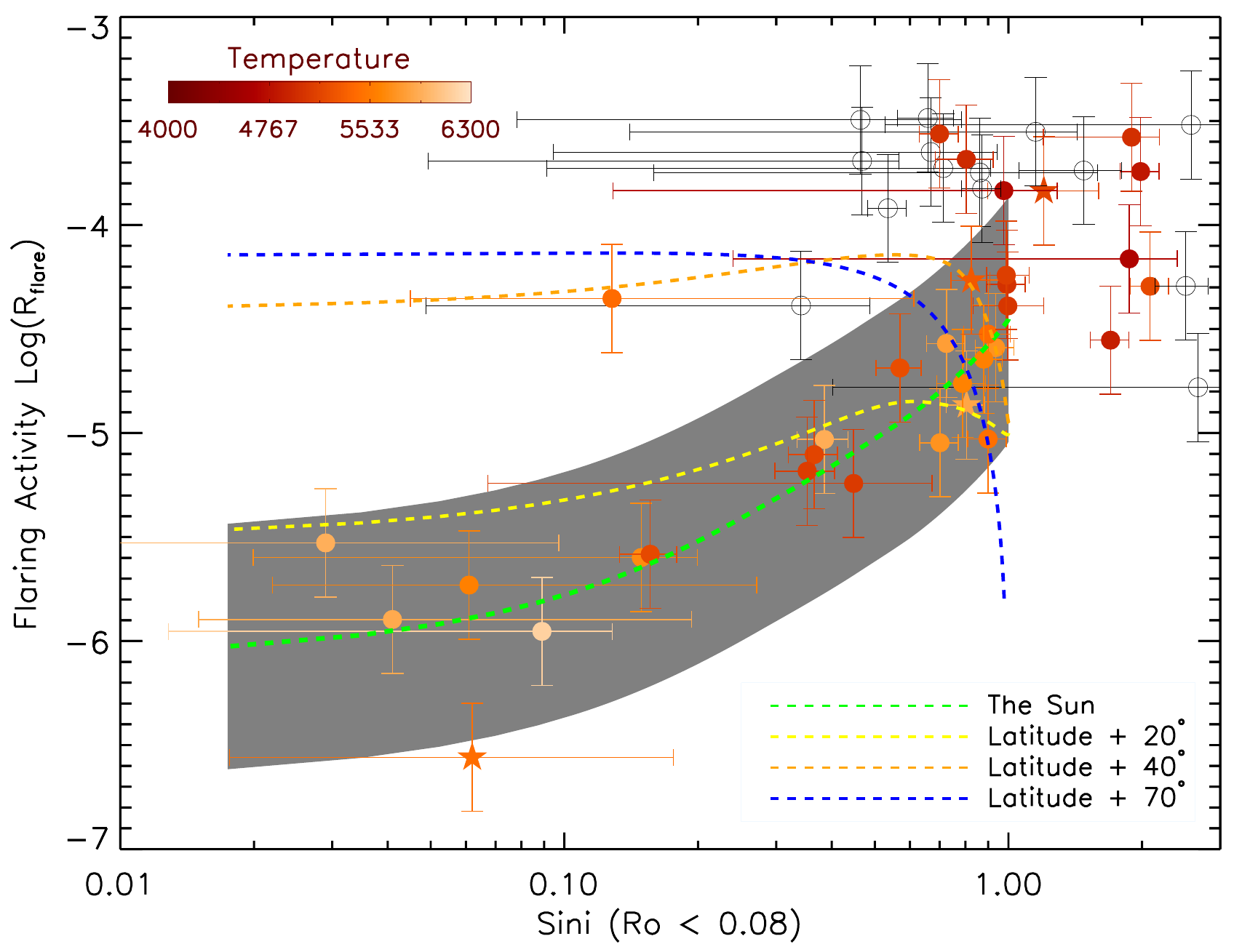}
\caption{Relation between flaring activity and inclination in the saturated regime (Ro$<$0.08). Circles denote dwarfs, and five-pointed stars denote subgiants. Open circles represent M-type stars. The green line denotes the variation in the solar flaring with the inclination and the shaded region represents the uncertainty caused by the maximum and minimum of the solar cycle. The other three dashed lines denote the variation caused by the mock data and correspond to the flaring region distribution at mid latitude(+20$^{\circ}$), high latitude (+40$^{\circ}$), and polar region (+70$^{\circ}$), respectively. All the dashed lines have been shifted vertically for a comparison. }
\label{fig3}
\end{figure}

\begin{figure*}
\includegraphics[width=0.32\textwidth]{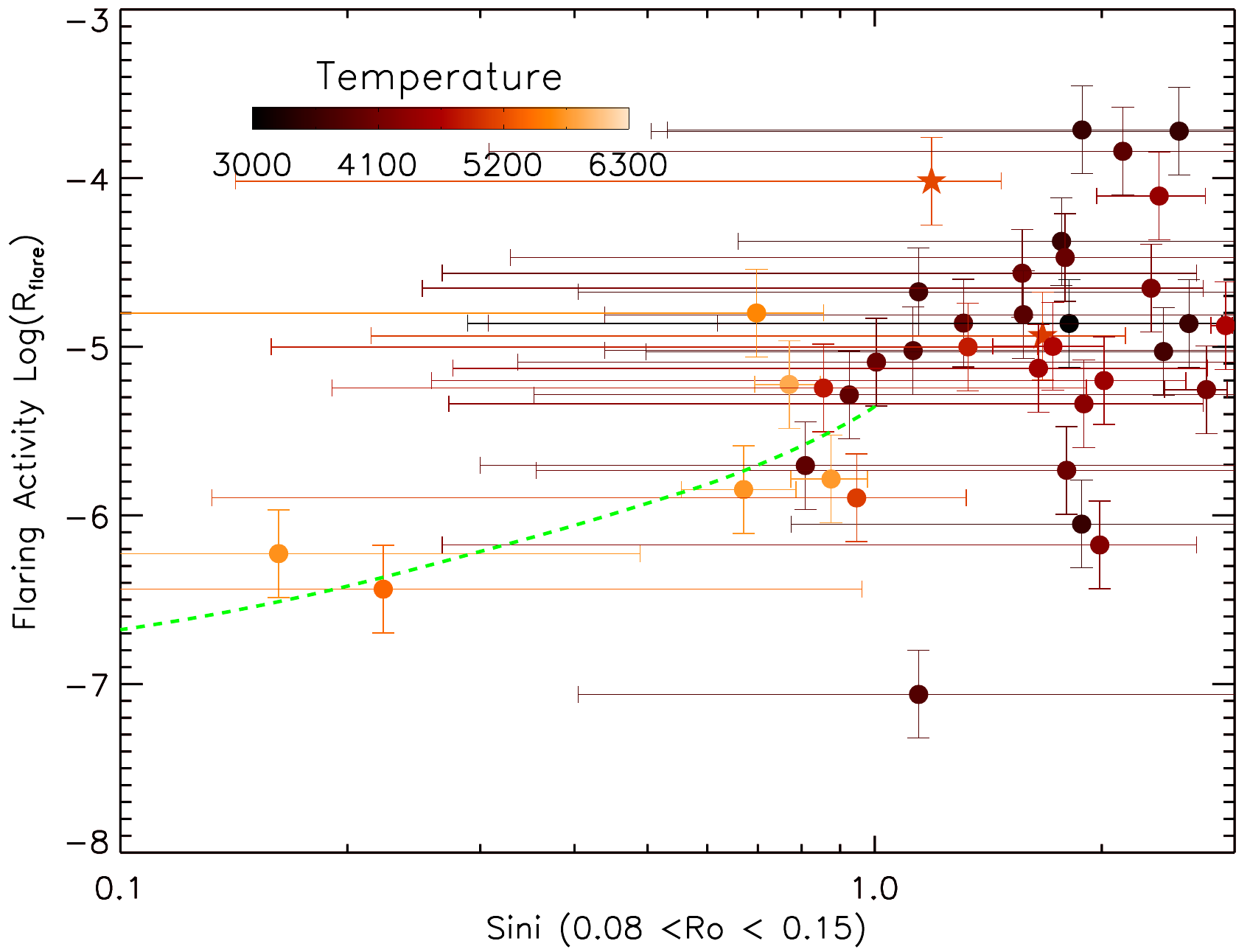}
\includegraphics[width=0.32\textwidth]{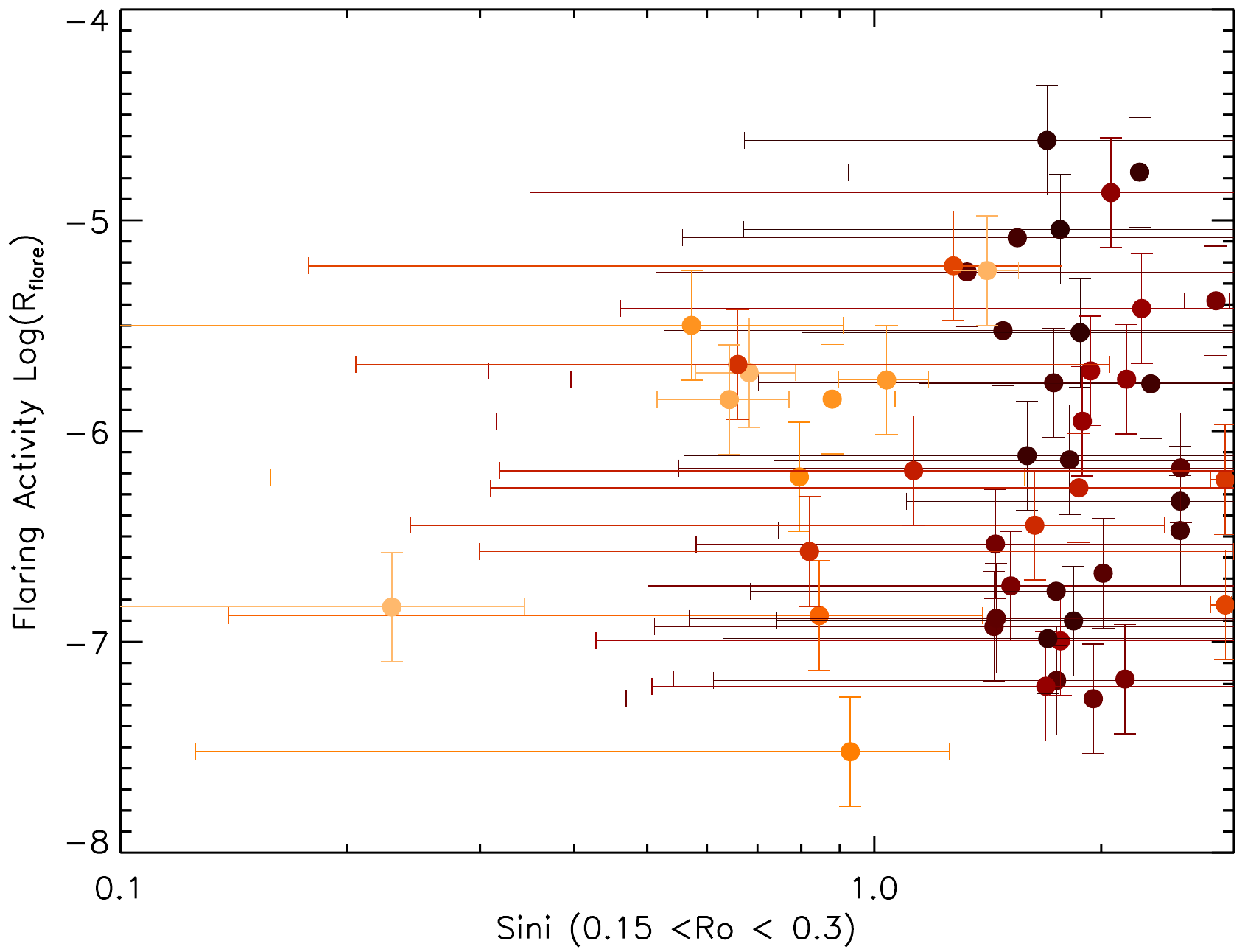}
\includegraphics[width=0.32\textwidth]{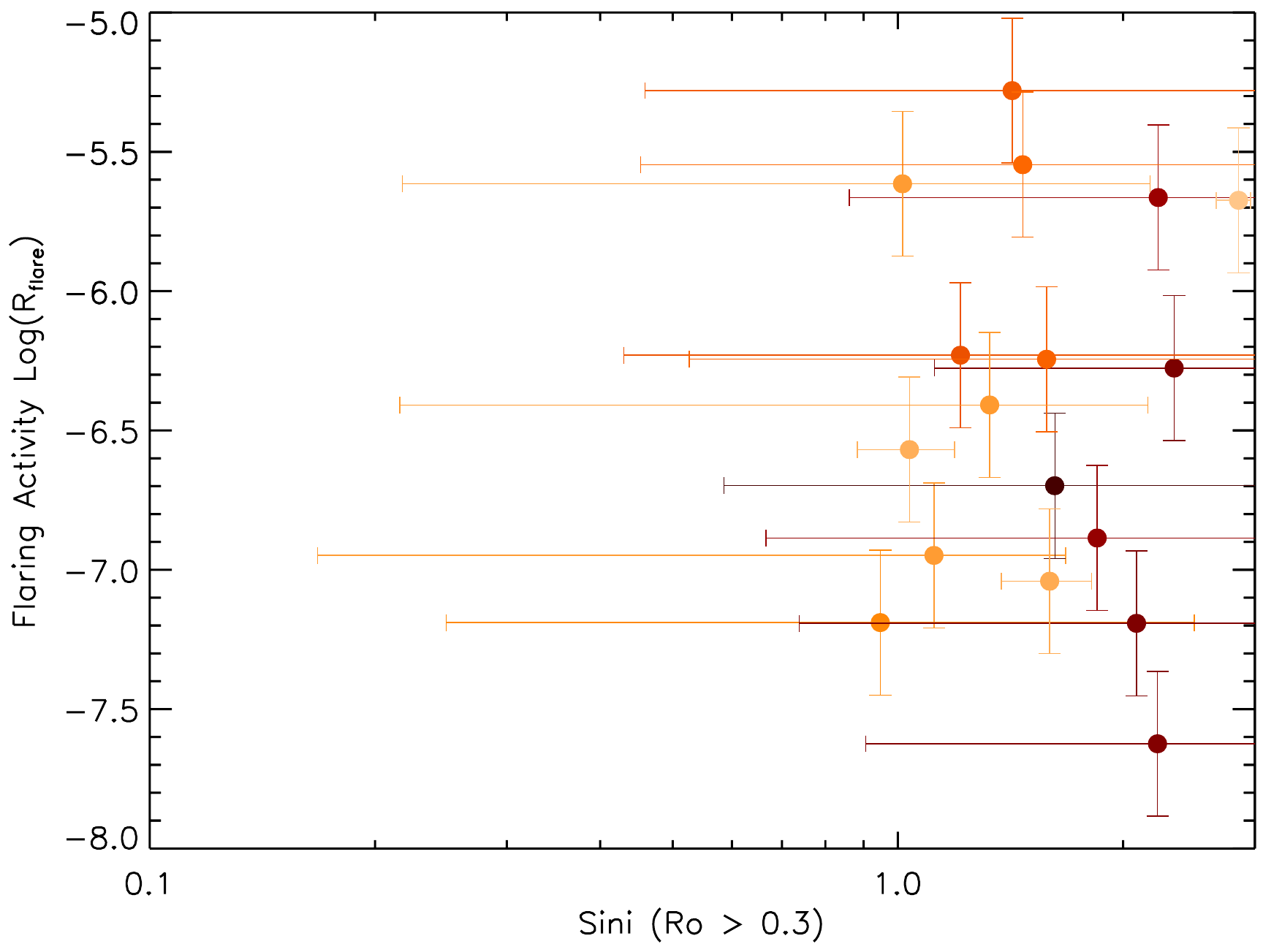}

\caption{Flaring activity vs. inclination for stars in the unsaturated regime. In order to reduce the influence of rotation, the Rossby number was divided into three intervals.  The large uncertainty is due to the decline in $v$sin$i$ with decreasing rotation and our conservative policy for stars of $v$sin$i < 10 {\rm ~km~s^{-1}}$ (their upper limit and lower limit \textcolor{red}{\st{of them}} were set to be $1{\rm ~km~s^{-1}}$ and $10{\rm ~km~s^{-1}}$ , respectively). }
\label{fig_unsate}
\end{figure*}

\begin{figure*}
\includegraphics[width=0.5\textwidth]{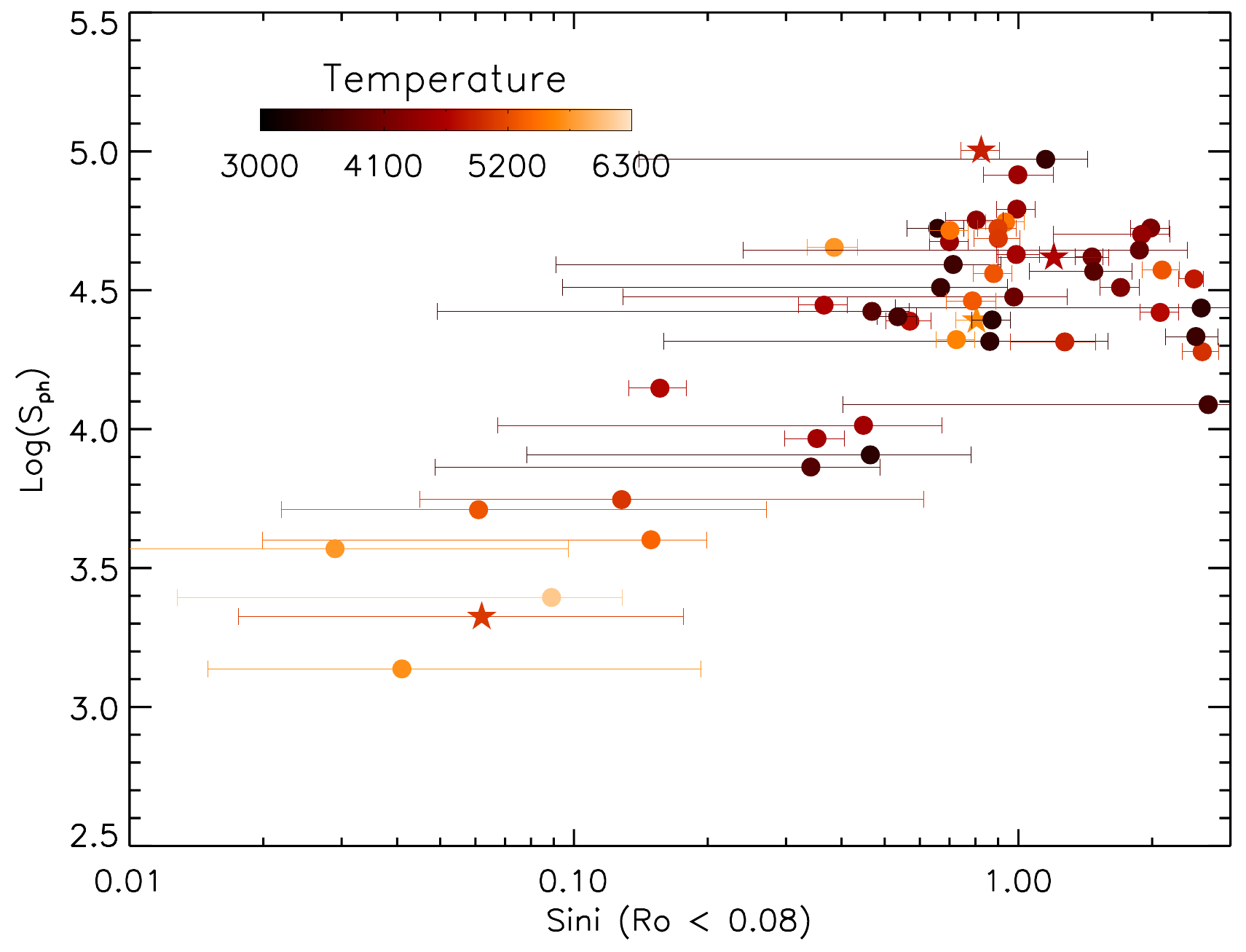}
\includegraphics[width=0.5\textwidth]{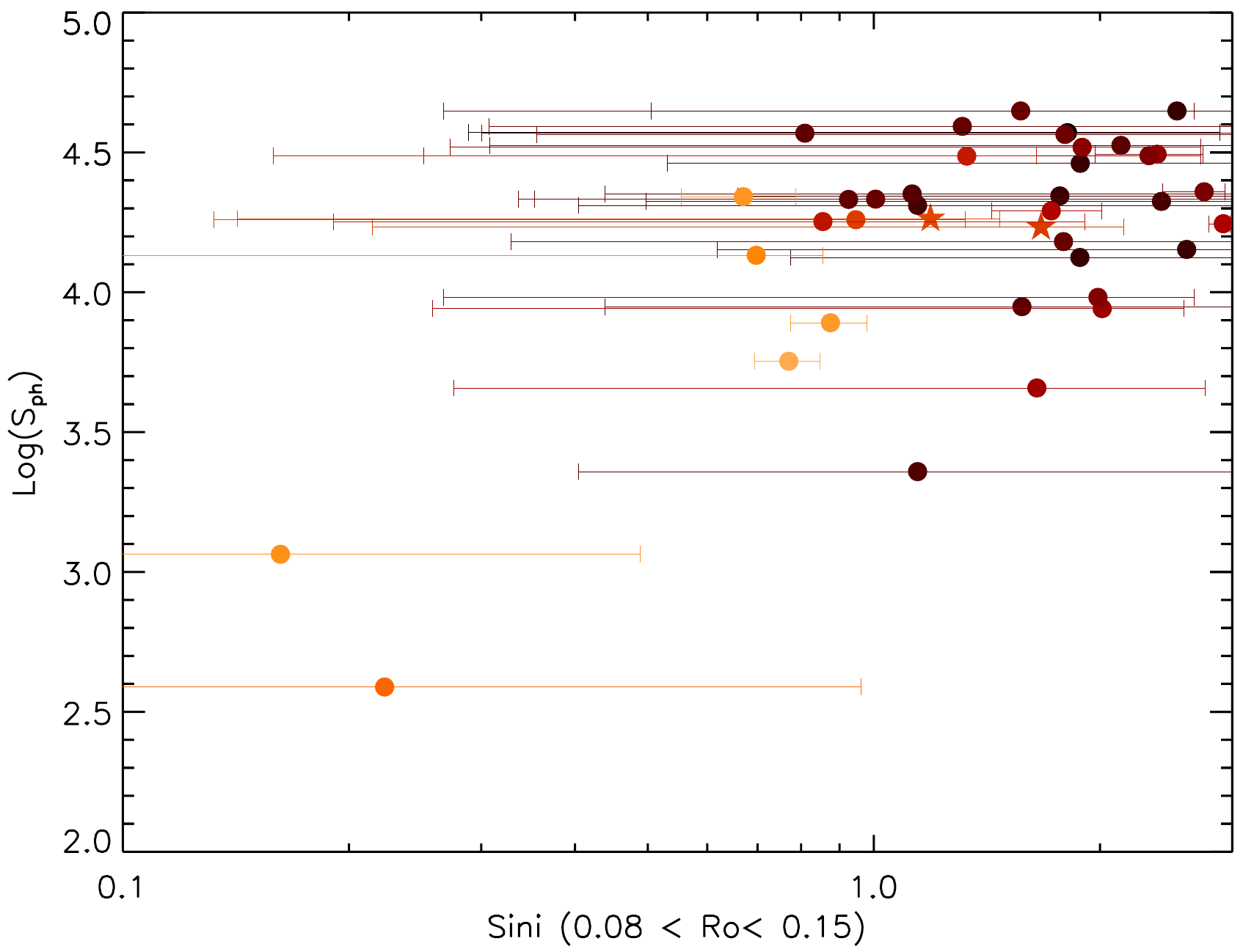}
\includegraphics[width=0.5\textwidth]{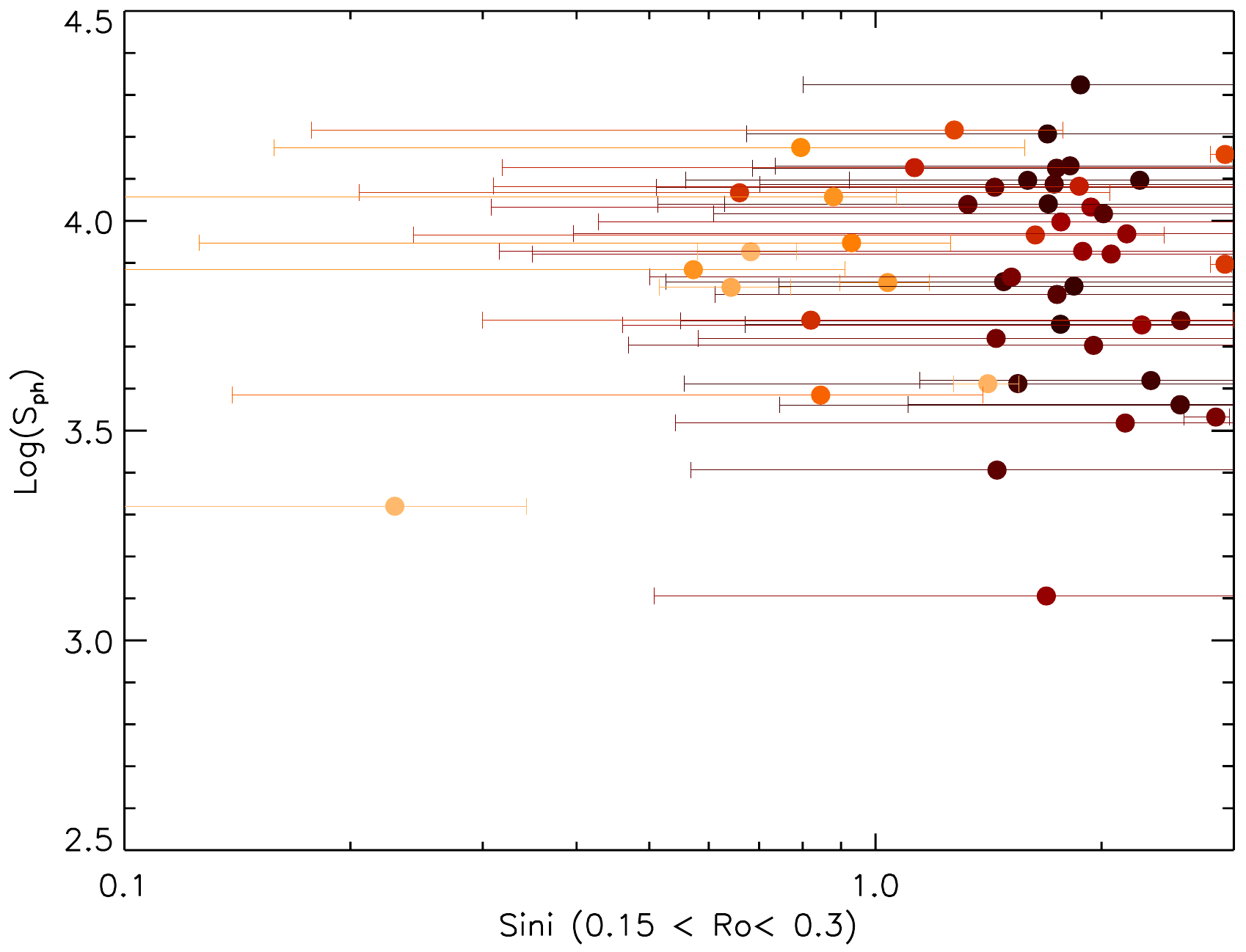}
\includegraphics[width=0.5\textwidth]{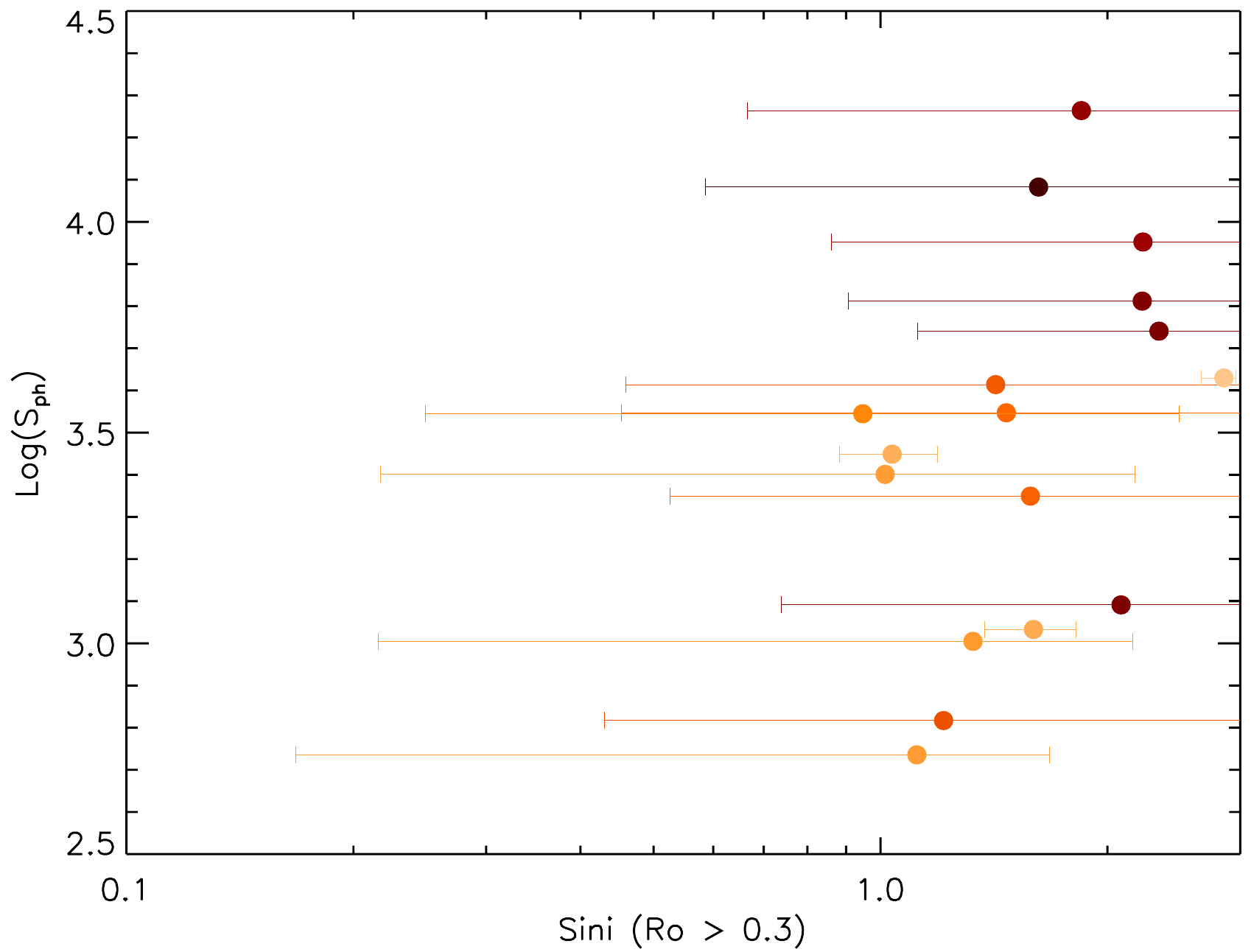}
\caption{Amplitude of light curve modulation $S_{\rm ph}$ \citep{MC2014} vs. inclination for a different range of Rossby number.}
\label{fig_sph}
\end{figure*}

\section{Result and discussion}
\subsection{Comparison between the Sun and stars on the latitudinal distributions of flaring activities}

We derived the inclinations of the flaring stars shown in Fig. 1 using three observational quantities: the projected rotational velocity ($v$sin$i$)  \citep{Abdu2022}, rotation period  \citep{MC2014}, and stellar radius  \citep{Somers2020}. The values of sin$i$ greater than 1 are for nearly equator-on stars or probably unresolved binaries  \citep{Healy2023,Simonian2020}. We compared the inclination sin$i$ and the apparent flaring activity in the saturated regime (Fig.~\ref{fig3}) and unsaturated regime (Fig.~\ref{fig_unsate}).

In the saturated regime, the magnetic field strength of fast rotating stars becomes saturated \citep{Reiners2009}. Their intrinsic activities are widely believed to be independent of rotation and should have the same activity level  \citep{Noyes1984,Wright2011,Yang2017}. However, we found that the variation in the apparent flaring activities we derived is related to the inclination (Fig.~\ref{fig3}). This most likely results from the LDAR of those fast rotating stars. The comparison between the stellar and solar variations with different latitude distributions of flares (dashed lines of Fig.~\ref{fig3}) shows that, for F-, G-, and K-type stars (4000 K$<T_{\rm eff}<$6300 K) with a fast rotation, their LDAR is solar-like. For M-type stars ($T_{\rm eff}<$4000 K), although they have a higher activity level, we do not have sufficient observations to identify their relationship with the inclination. 

In the unsaturated regime, we divided the Rossby number into three intervals to reduce the influence of rotation (Fig.~\ref{fig_unsate}). In each interval, we assumed that the stars have the same intrinsic flaring activity and checked whether they conform to the solar relation. The uncertainty of $v$sin$i$ increases rapidly as stars slow down and it is even impossible to detect flaring stars with low inclination because of the exponential decay of the apparent flaring activity along with decreasing inclination. In spite of that, we still find a trend (or a positive correlation) between the apparent flaring activity and inclination, especially for relatively fast rotators (0.08<Ro<0.15 the top left panel). As the rotation decreases, this trend becomes more and more ambiguous because of the observation limit. Given that stars in the unsaturated regime are supposed to be solar-like dynamo, they are more likely to have the same relation as the Sun. We suggest that better observations may validate this trend in the future.

\subsection{Interpretation of the LDAR of fast rotating stars}

On the Sun, very few flares were found outside of active regions \citep{Hath2015}. The latitudinal distribution of flares can represent the LDAR. The similarity of flaring latitudes demonstrates that the LDAR of fast rotating stars is highly similar to that of the Sun (Fig.~\ref{fig3}). 

Various tests have shown that Doppler imaging is inefficient for the equatorial region, where spots are recovered with reduced area and contrast  \citep{Berdy2005,Rice2002}. Our results find crucial evidence for flares being mainly distributed at latitudes of $\theta \leq 30^{\circ}$. Moreover, our sample includes about 50 stars with over four years of observations. Since a faster rotator has a shorter stellar cycle \citep{Boro2018}, the four-year observations may cover the whole stellar cycle of those stars. If their starspots propagate poleward, in order to ensure that starspots are always at low latitudes of $\theta \leq 30^{\circ}$ throughout a whole stellar cycle, their corresponding initial emergence latitudes should be at latitudes of $\theta \leq 10^{\circ}$. However, fast rotating stars tend to form initial starspots at higher latitudes of $\theta \geq 30^{\circ}$ where the latitudinal differential rotation is stronger and the toroidal field generation is more efficient  \citep{Isik2018,Zhang2024}. Therefore, we suggest that starspots of fast rotating stars begin to appear at a latitude of around $30^{\circ}$  and propagate equatorward, closely resembling what happens on the Sun.

Another interesting and still controversial feature in stellar Doppler images are large polar spots or cool active regions over the stellar poles, which are the so-called polar cap  \citep{Strass2002,Berdy2005}. This feature is explained by the scenario that faster rotators give rise to higher latitudes of active region  \citep{Strass2002,Berdy2005}. Our results do not support the scenario of a polar cap for F-, G-, and K-type stars because this would result in a totally different relation between the flaring activity and inclination (the dashed blue line in Fig.~\ref{fig3}).

Admittedly, the prevailing viewpoint based on Doppler imaging and simulations postulates that polar spots are widespread \citep[e.g.,][]{Schuessler1992,Choudhuri1987,Strass2002}, which contradicts our results. We should note that  polar caps could be artifacts caused by stellar activity \citep{Bruls1998}, microturbulence and blend lines \citep{Unruh1995}), the uncertainty of vsini \citep{Berdy1998}, or  differential rotation \citep{Hackman2001}, although these factors are not a general explanation for all polar caps. Moreover, the capacity of Doppler imaging to recover features near the equator is seriously weakened because of the less differential rotation, reduced area, and contrast \citep{Berdy1998,Berdy2005,Rice2002}. Features near the equator are smeared and only the strongest features can be recovered \citep{Rice1989,Rice2002}. These disadvantages of Doppler imaging indicate that our result of LDAR near the equator could be more reliable. Another potential explanation for the contradiction is that polar spots are more stable (or eruption is difficult), such that flares at the polar cap are rare. This explanation may be supported by  \citet{Younes2020}, who found that a flare of the polar cap was triggered by the global dipolar pattern, which implies that eruption is difficult there. However, we suggest that, to settle the contradiction, further studies on flares of those famous polar-cap stars could provide critical information.

Previous studies on fast rotating M dwarfs found that their magnetic field geometries had two states: dipole and multipole  \citep{Gastine2013,Koch2021}. The magnetic field strength of the multipole state saturates at about 4 kG, while that of the dipole state could exceed 4 kG and does not exhibit a saturation  \citep{Shulyak2017}. The difference in the magnetic field strength of the two states was also revealed by the activity proxy of X-ray luminosity \citep{Shulyak2017}. The dipole state of M dwarfs may account for their higher activity level, as shown in Fig.~\ref{fig3}. However, the different activity levels of the two states make it difficult to determine the relationship between the activity and inclination for M-type stars. Nevertheless, given that M dwarfs in a dipole state are associated with the polar cap \citep{Morin2010} and that giant flares have been found at high latitudes of fully convective stars  \citep{Ilin2021}, we suggest that M-type stars may have a different LDAR from the Sun.

The uncertainty of the inclination of M-type stars increases rapidly as $v$sin$i$ declines with decreasing stellar mass. Their low luminosity also renders them less likely to be observed with high-resolution spectroscopy and high-precision photometry. To verify the relationship between the flaring activity and inclination for M dwarfs, more high-precision observations are required.

\subsection{Can the LDAR be revealed by other proxies?}
Flaring activity is probably the only proxy that can reveal the LDAR of stars under the current conditions of observation because: (i) It only reflects the contribution from active regions. The chromospheric activity ($R'_{\rm HK}$ ) and coronal activity ($R_{\rm X}$ ) include the contribution from the whole stellar hemisphere (regardless of whether there is an active region or not, a local stellar region will make a contribution to the activity proxy $R'_{\rm HK}$ and $R_{\rm X}$). This may greatly dilute the contribution of active regions, with the result that the variation in those proxies is small. For example, the variation in $R'_{\rm HK}$ between the maximum and minimum of a solar cycle is $\sim 18\%$ \citep{Egeland2017}, while the variation in flaring activity is at least $360\%$ (Fig.~\ref{fig_circle}). We investigated the relation between $R'_{\rm HK}$ and inclination and did not find any solid relation. (ii) The observation of a flare is independent of its inclination. For example, the proxy of light curve modulation  \citep{MC2014,Santos2021} caused by starspots only reflects the contribution from active regions. However, its amplitude $(S_{\rm ph}$ ) depends on the inclination. It will be smaller if observed from pole-on and larger from equator-on (Fig.~\ref{fig_sph}), as the invisible region decreases. In other words, from
Fig.~ \ref{fig_sph} we cannot identify whether the variation in the amplitude is caused by the LDAR or inclination. However, the top left panel of Fig. ~\ref{fig_sph}  validates the calculation of inclination for fast rotating stars (lower inclination should have smaller $S_{\rm ph}$) and indicates that $S_{\rm ph}$ and the flaring activity have a strong correlation.

\subsection{Implication for the dynamo theory}


In the dynamo simulation of the Sun, the most important feature that needs to be reproduced is that sunspots appear at low latitudes and propagate equatorward  \citep{Charb2020,Nandy2002}. The LDAR provides direct and strong evidence of the underlying dynamo process  \citep{Roett2016}. In our sample, the similarity in the LDAR between fast rotating stars and the Sun demonstrates that they may have the same dynamo process, thus providing a crucial constraint to the stellar dynamo theory and simulation. 

Our results are contrary to the expectation raised by previous dynamo simulations \citep{Isik2018,Zhang2024}. They predict that faster rotators have higher latitudes of active regions and form an inactive gap near the equator because of the weak latitudinal shear and the low efficiency of toroidal field generation near the equator. To settle this discrepancy, the distribution of the latitudinal differential rotation and the role of the radial shear need to be reconsidered \citep{Nelson2013}. 

Regarding the wider scope of the dynamo theory, the popular viewpoint on the stellar dynamo is that stars in the saturated and unsaturated regime have different dynamos (convective and solar-like dynamos), which are based on the activity dependence on rotation \citep{Noyes1984,Wright2011,Wright2016,Barnes2007}.  Our result challenges this viewpoint because the convective and solar-like dynamo give rise to similar results, as found here. If this is correct, it necessitates the establishment of a unified solar-like dynamo that spans the whole evolutionary stage of a star. 

The study by \citet{Wright2016} also found that fully convective stars had the same activity dependence on rotation as partially convective stars, indicating that they have a solar-like dynamo  . Since fully convective stars do not have a tachocline, it questions the canonical scenario that a tachocline is crucial for the solar dynamo in the unsaturated regime. Because our stars are in the saturated regime, which includes very young main sequence or pre-main sequence stars \citep{Barnes2007}, they do not have a tachocline either, probably because the radiative core and convective envelop have not yet been coupled \citep{Brun2017,Barnes2007}. Therefore our results further challenge the importance of the tachocline in the solar-like dynamo, supporting the theory proposed by recent dynamo simulations that the tachocline is unnecessary \citep{Charb2020,Nelson2013,Fan2014,Zhang2022}.




\section{Conculsion}
In this study, we calculated the flaring activity and inclination of about 200 stars using the 4-year light curves of the {\it Kepler} mission and the APOGEE spectra. We also used about 40 years of solar flaring data to simulate observations of the Sun by the {\it Kepler} mission from equator-on to pole-on, and we obtained the relationship between the apparent flaring activity of the Sun and inclination. We compared the relationship with flaring stars with different inclinations and found that fast rotators (Ro$<$0.08) have a similar relation to that of the Sun, which indicates that their LDAR are solar-like, while the relation of slow rotators (Ro$>$0.08) needs to be verified in the future due to the data quality and observation limitations.

Our results provide a crucial constraint to dynamo theory and its simulation, and also imply that fast rotators have the same dynamo process as the Sun, contrary to the numerical simulations. Meanwhile, our results strongly imply that a solar-like dynamo may be applied at all evolutionary stages of a star, which is also contrary to the popular viewpoint.

Moreover, our results contradict the prevailing viewpoint that polar spots are widespread in fast rotators. Although we present several potential reasons for the contradiction, we cannot obtain a solid and general explanation. We suggest that, to settle this issue, further studies on flares of those famous polar-cap stars could provide critical information.
\section*{Data availability}
Data used in this study are only available in electronic form at the CDS via anonymous ftp to cdsarc.u-strasbg.fr (130.79.128.5) or via http://cdsweb.u-strasbg.fr/cgi-bin/qcat?J/A+A/.
\begin{acknowledgements}
We sincerely thank J. Jiang, B.F. Healy, F. Spada and K. Namekata for discussion and help. Jifeng Liu acknowledges support from the New Cornerstone Science Foundation through the New Cornerstone Investigator Program and the XPLORER PRIZE. Xin Cheng and Zhanhao Zhao acknowledge support from the Strategic Priority Research Program of the Chinese Academy of Sciences, Grant No. XDB0560000. Yijun Hou acknowledges support from the National Natural Science Foundation of China (12273060), the Youth Innovation Promotion Association CAS (2023063) and the Strategic Priority Research Program of CAS (XDB0560000). ZeXi Niu acknowledges support from the National Natural Science Foundation of China through grant No. 12303039 and No. 12261141690.
\end{acknowledgements}

%
%
\bibliographystyle{aa}
\bibliography{aabib}
\onecolumn
\begin{appendix}
\section{Additional figures}

\begin{figure*}[h!]%
\includegraphics[width=1\textwidth,height=0.15\textheight]{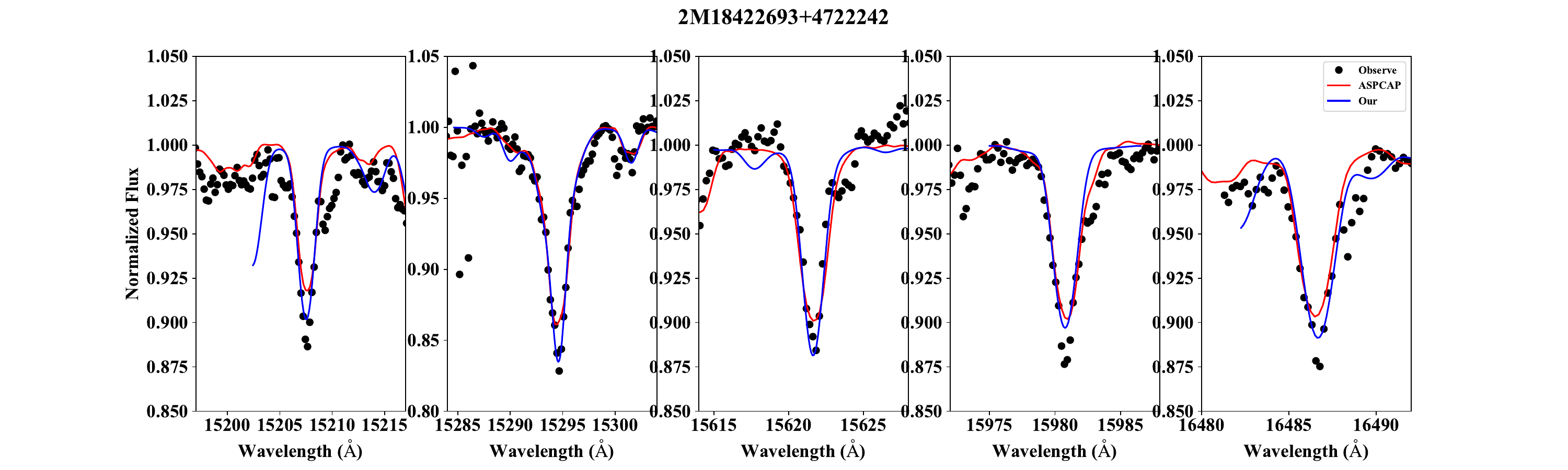}
\includegraphics[width=1\textwidth,height=0.15\textheight]{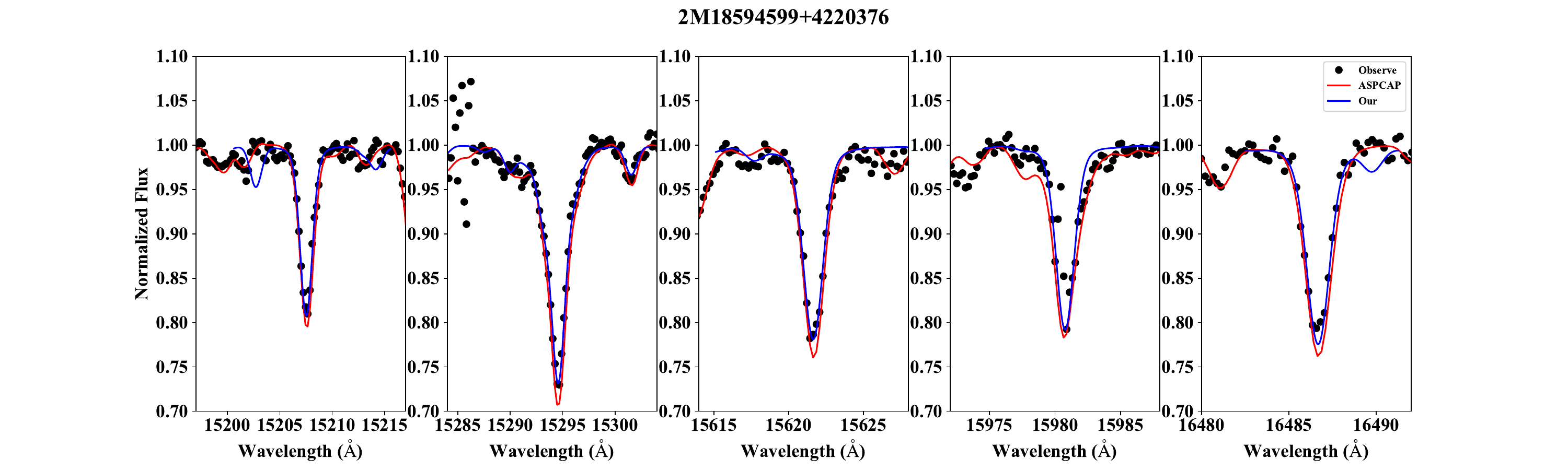}
\includegraphics[width=1\textwidth,height=0.15\textheight]{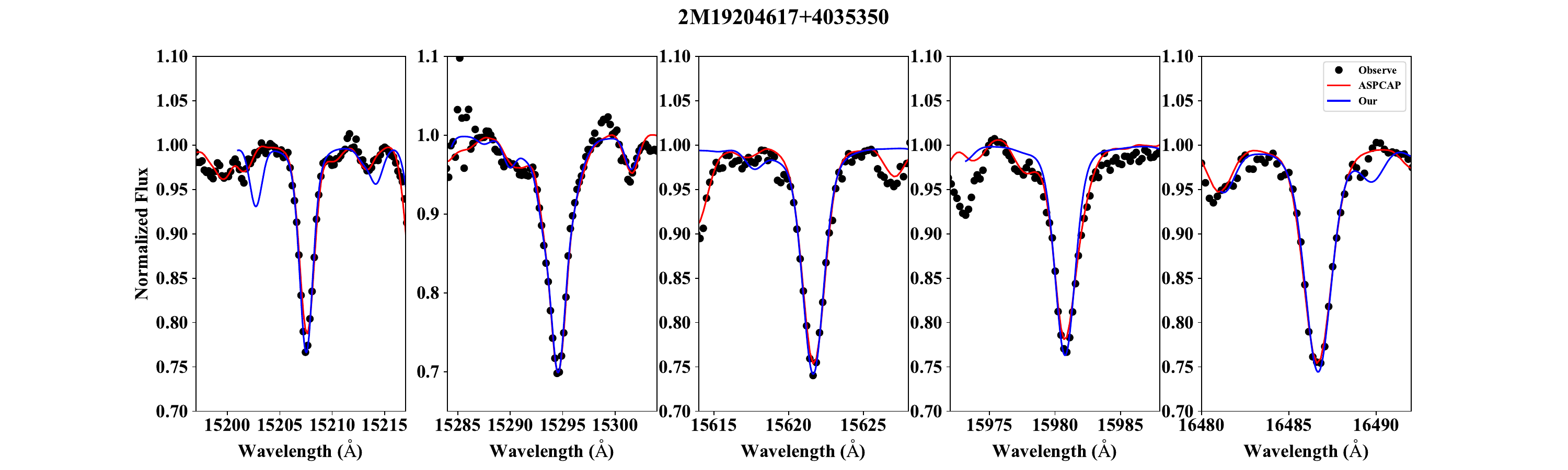}
\includegraphics[width=1\textwidth,height=0.15\textheight]{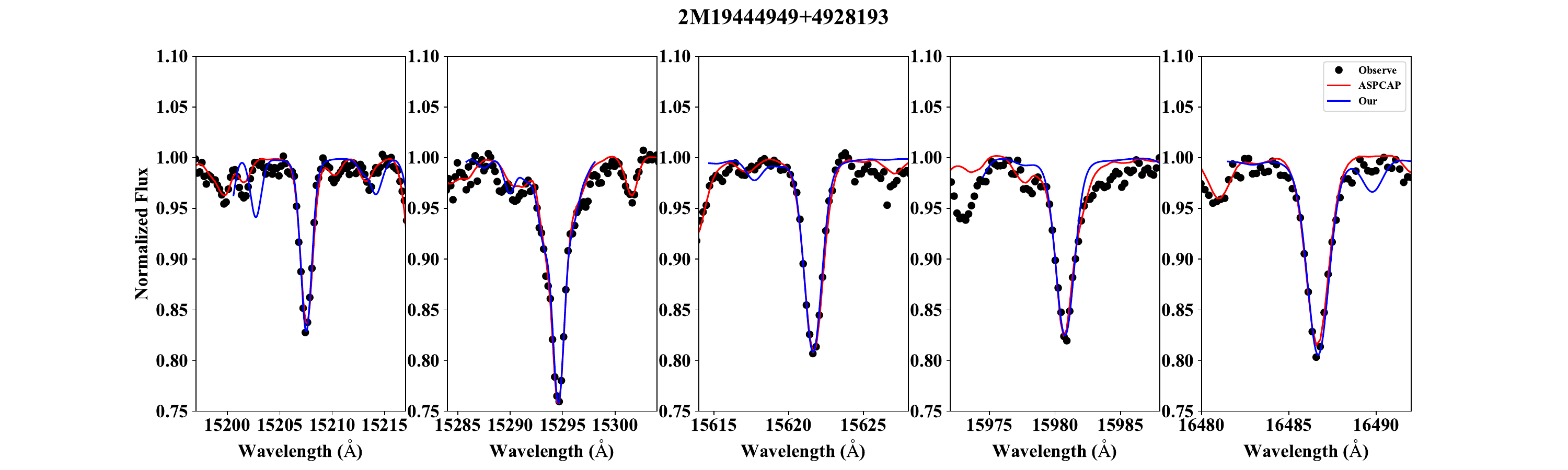}
\includegraphics[width=1\textwidth,height=0.15\textheight]{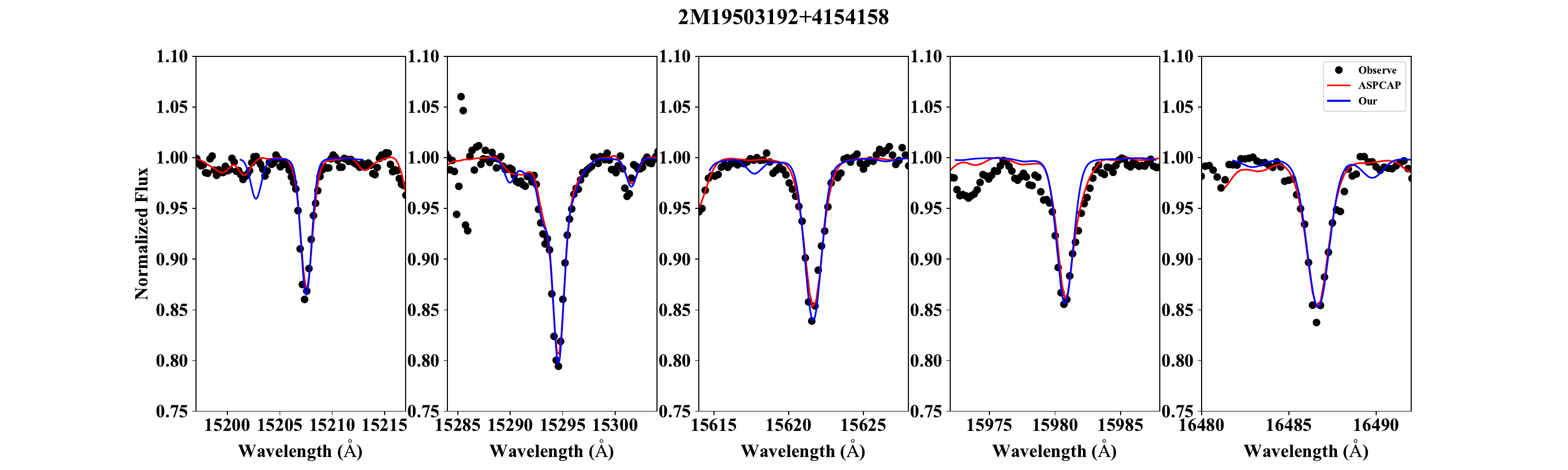}
\caption{Comparison of the spectrum fitting between our method and that of ASPCAP. Five stars whose difference of $v$sin$i$ is larger than $5{\rm ~km~s^{-1}}$ are plotted. Each row shows a star including five Fe I lines that exhibit minimal blending. Blue lines represent the our best-fit. Red lines represent the best-fit of ASPCAP.
}\label{fig_vsini_examp2}
\end{figure*}
\begin{figure*}%
\includegraphics[width=1\textwidth,height=0.15\textheight]{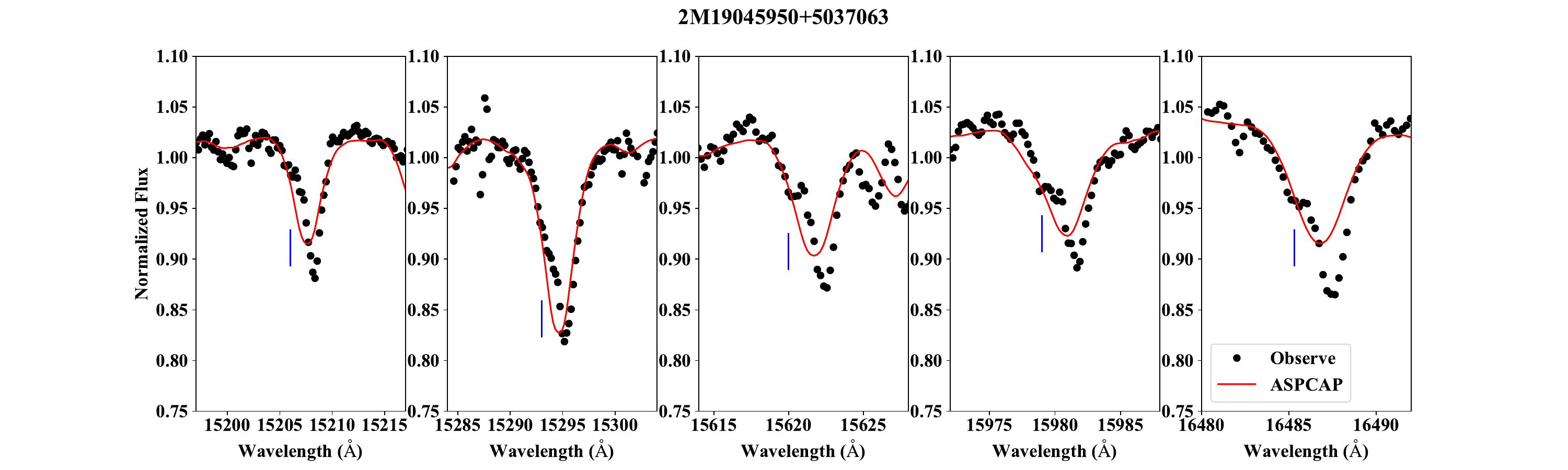}
\includegraphics[width=1\textwidth,height=0.15\textheight]{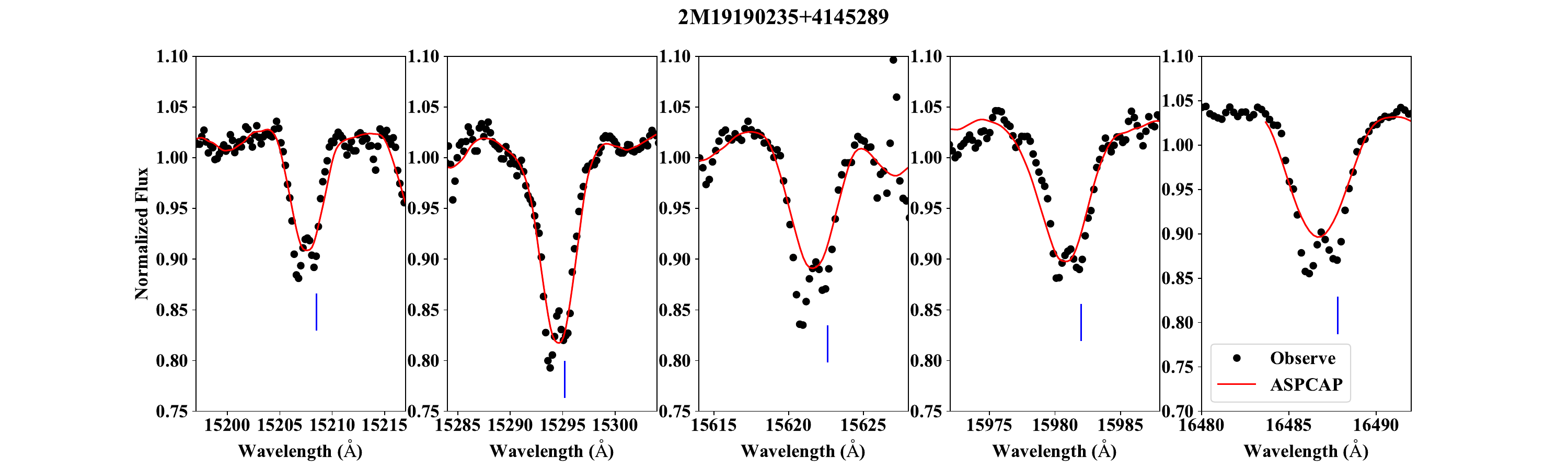}
\includegraphics[width=1\textwidth,height=0.15\textheight]{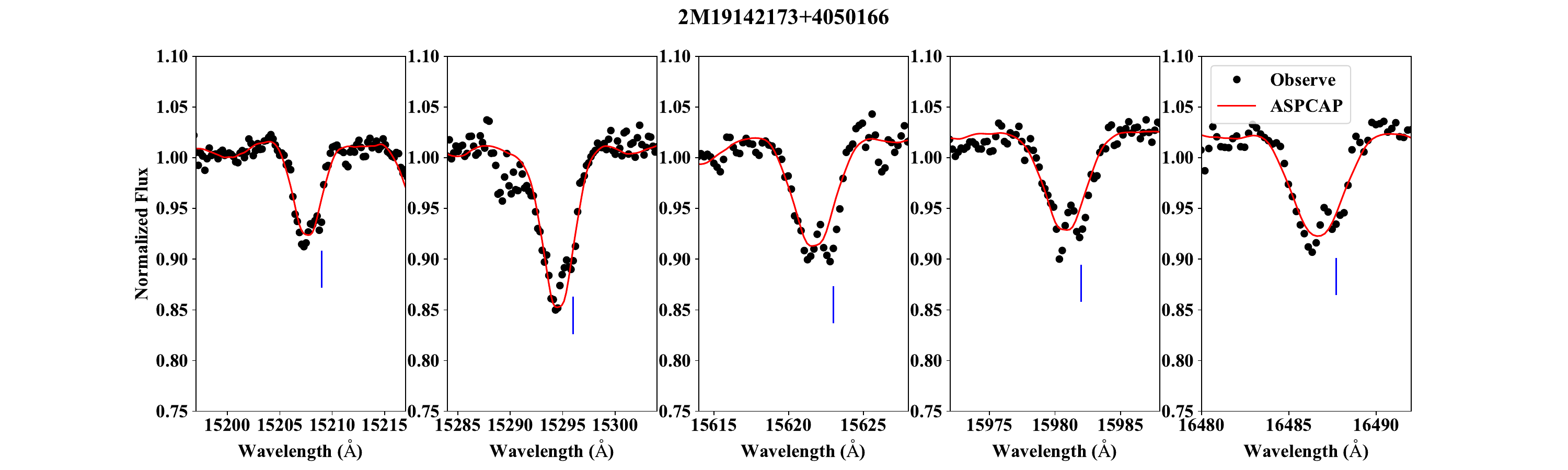}
\includegraphics[width=1\textwidth,height=0.15\textheight]{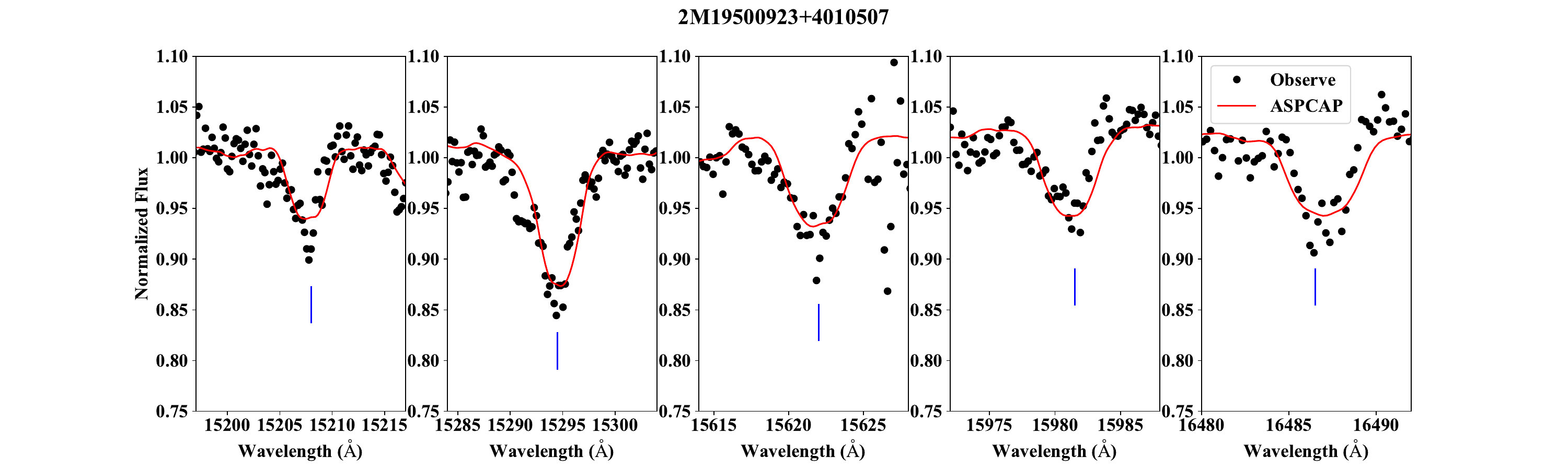}
\includegraphics[width=1\textwidth,height=0.15\textheight]{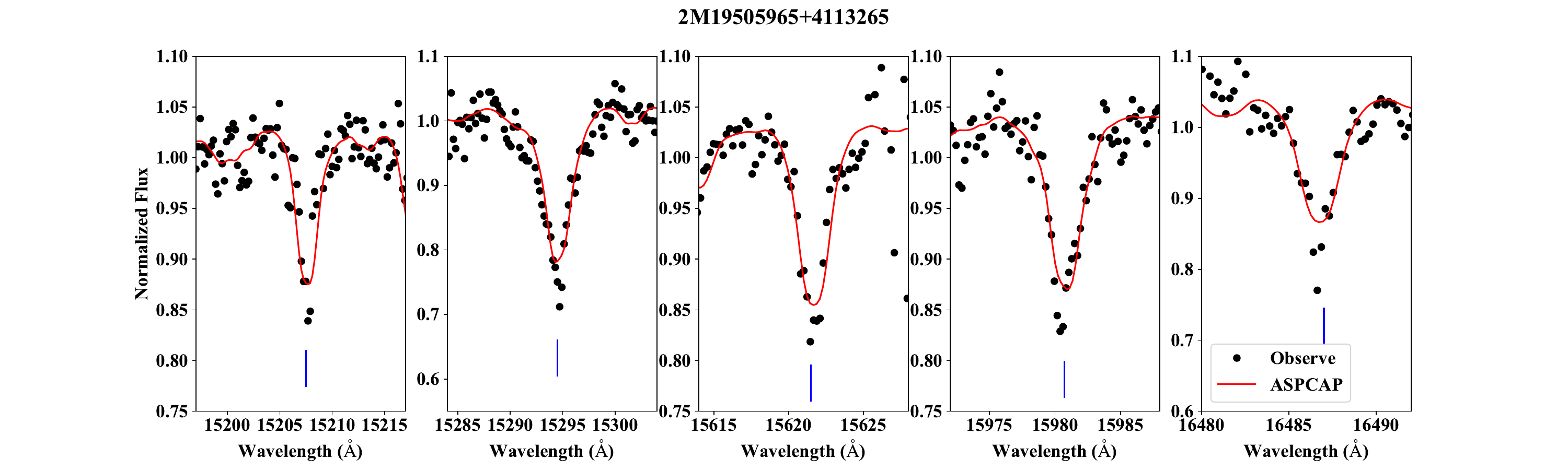}
\includegraphics[width=1\textwidth,height=0.15\textheight]{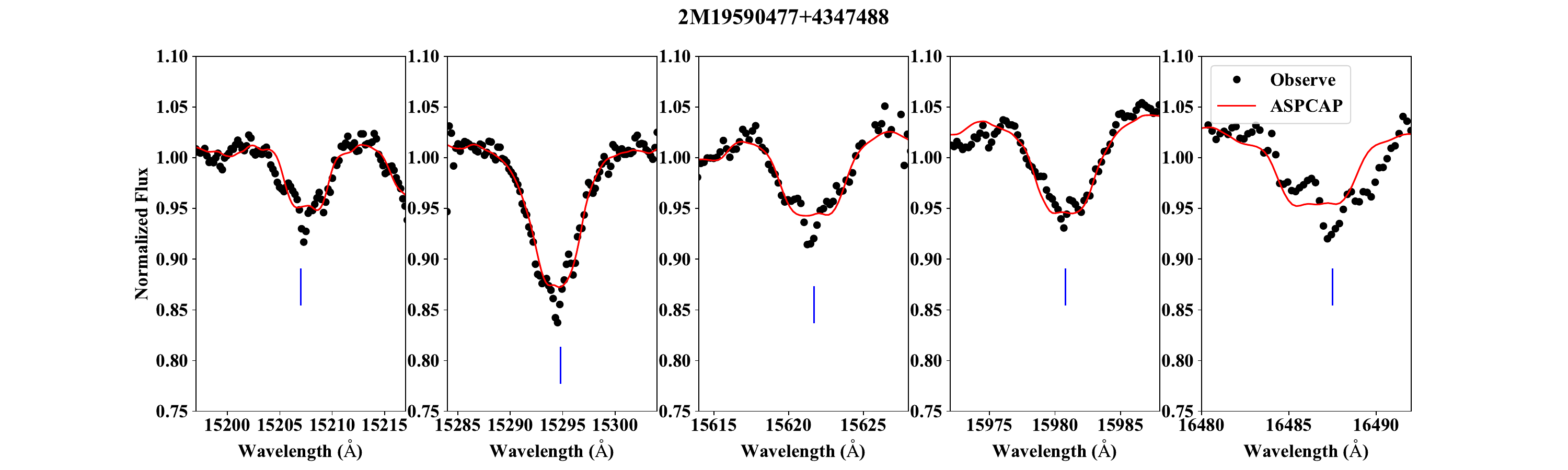}
\caption{Illustrated examples of six visual SB2s identified in our sample. Each row shows a star including five Fe I lines that exhibit minimal blending. The vertical blue lines denote the possible companion feature. Red line represent the best-fit of ASPCAP. Those six binaries are plotted with triangle in Fig. 1, but are not included in the further analysis. 
}\label{fig_vsini_examp}
\end{figure*}
\end{appendix}

\end{document}